\documentclass[journal=jacsat,manuscript=article]{achemso}
\usepackage[version=3]{mhchem} 
\usepackage{color}
\usepackage{multirow}
\usepackage{float}
\usepackage{hyperref}
\usepackage{lineno}
\usepackage{amsmath}
\usepackage{amsfonts}

\usepackage[scr=boondox,  
            cal=esstix]   
           {mathalpha}

\newcommand{\oo}{$\mathcal{O}$ }
\newcommand{\qq}{$\textbf{Q}$ }
\newcommand{\dd}{$\textbf{D}$ }
\newcommand{\pp}{$\textbf{P}$ }
\newcommand{\rr}{$\textbf{R}$ }
\newcommand{\sq}{$\mathcal{q}$ }
\newcommand{\sm}{$\mathcal{m}$ }
\usepackage{pdfpages}

\usepackage{etoolbox}
\makeatletter
\renewcommand*{\acs@author@fnsymbol@symbol}[1]{
    \ifcase #1 *\or
    1\or
    2\or
    3\or
    4\or
    5\or
    6\or
    7\or
    8\or
    9\or
    10
    \fi
}
\makeatletter
\patchcmd{\acs@contact@details}{E}{*\,E}{}{}
\makeatother
\author{Akash Pandey}
\author{Wei Chen}
\affiliation[1]
{Department of Mechanical Engineering, Northwestern University, Evanston, Illinois 60208, United States}
\author{Sinan Keten}
\email{s-keten@northwestern.edu}
\affiliation[1]
{Department of Mechanical Engineering, Northwestern University, Evanston, Illinois 60208, United States}
\alsoaffiliation[2]
{Department of Civil and Environmental Engineering, Northwestern University, Evanston,  Illinois 60208, United States}
\title[An \textsf{achemso} demo]
  {COLOR: A compositional linear operation-based representation of protein sequences for identification of monomer contributions to properties}
\abbreviations{IR,NMR,UV}

\begin{document}

\begin{abstract}

The properties of biological materials like proteins and nucleic acids are largely determined by their primary sequence. While certain segments in the sequence strongly influence specific functions, identifying these segments, or so-called motifs, is challenging due to the complexity of sequential data. While deep learning (DL) models can accurately capture sequence-property relationships, the degree of nonlinearity in these models limits the assessment of monomer contributions to a property - a critical step in identifying key motifs. Recent advances in explainable AI (XAI) offer attention and gradient-based methods for estimating monomeric contributions. However, these methods are primarily applied to classification tasks, such as binding site identification, where they achieve limited accuracy (40–45\%) and rely on qualitative evaluations. To address these limitations, we introduce a DL model with interpretable steps, enabling direct tracing of monomeric contributions. We also propose a metric ($\mathcal{I}$), inspired by the masking technique in the field of image analysis and natural language processing, for quantitative analysis on datasets mainly containing distinct properties of anti-cancer peptides (ACP), antimicrobial peptides (AMP), and collagen. Our model exhibits 22\% higher explainability, pinpoints critical motifs (RRR, RRI, and RSS) that significantly destabilize ACPs, and identifies motifs in AMPs that are 50\% more effective in converting non-AMPs to AMPs. These findings highlight the potential of our model in guiding mutation strategies for designing  protein-based biomaterials.



\end{abstract}
\section{Main}

Proteins are defined by their amino acid sequences, also known as the primary sequence, which dictates their structure and function \cite{branden2012introduction}. Specific residues or segments within the sequence often play pivotal roles in functionality \cite{arakawa20221000, QuoteTarget, konc2022protein}, but identifying these critical regions from just the primary sequence remains a challenge as the residue names represent dense chemical information. Numerous studies have explored the relationship between primary sequence and protein function using experimental techniques like X-ray crystallography, solid-state NMR, and Raman spectroscopy \cite{experiment1, experiment2, experiment3, experiment4, experiment5, md_mech}, as well as computational approaches like Molecular Dynamics (MD) simulations \cite{md1, md2, md5_mech, md6}. Moreover, MD simulations offer mechanistic insights into the fundamental processes such as the effect of the processing conditions on the mechanical properties \cite{md3_mech, md4_mech,md_mech}, transport mechanism of proteins across cell membrane \cite{md7_mech}, and the mechanism of protein folding \cite{karplus2005molecular}. Both experimental techniques \cite{monomers_are_preferred, li2023bi, eijsink2005directed, counago2006vivo, nemtseva2019experimental} and MD simulations \cite{sars_cov2, mou2015using, graham2023increase} have been used to investigate the impact of mutations on functionalities like binding affinity and mechanical properties. However, their time-intensive nature limits the ability to exhaustively explore mutations across the sequence, highlighting the need for predictive models that can efficiently estimate the contribution of individual monomers to overall protein function.

Machine learning (ML) models have emerged as a powerful tool for establishing primary sequence-to-property relationships in proteins \cite{brandes2022proteinbert, xu2020deep, prottrans}. This remains an active research area, as sequence data is more accessible than structural data. Deep learning (DL) models like AlphaFold2/3 \cite{alphafold, alphafold3} predict structures from sequences but often show low confidence for amorphous and fold-switching proteins \cite{alphafoldfail, alphafoldfail2}. This limits their reliability for materials with more disordered regions, notably structural proteins such as silks \cite{lefevre2007protein}. Models like Transformers \cite{vaswani2017attention}, Long Short-Term Memory (LSTM) networks \cite{hochreiter1997long,lstmreview,mohapatra2023person,mohapatra2023effect}, and 1D convolution Neural Networks (1D CNN) \cite{kiranyaz20191, mohapatra2024phase, cnnreview} excel at capturing sequential dependencies. These models enable accurate prediction of protein's secondary structure \cite{yu2022end}, antimicrobial capability \cite{gupta2019feedback}, B-factor \cite{pandey2023b, sun2019utility}, and mechanical properties \cite{liu2022presto}, often achieving $R^2$ and accuracies over 0.8. Transformers have also enabled pre-trained models like ProtBERT \cite{brandes2022proteinbert, prottrans}, ESM \cite{lin2023evolutionary}, and ProtTXL \cite{prottrans}, which differ in training strategies. For example, ProtTXL uses an auto-regressive method, while ProtBERT predicts masked monomers \cite{prottrans}. Trained on millions of protein sequences \cite{boutet2016uniprotkb}, these models capture patterns like sequential relationships \cite{detlefsen2022learning, brandes2022proteinbert}, and clustering based on protein families and physicochemical properties \cite{detlefsen2022learning, prottrans}. Their success has driven transfer learning frameworks, leveraging pre-trained model outputs as inputs to neural networks for predicting protein properties \cite{lin2023evolutionary, brandes2022proteinbert, khare2022collagentransformer}. Despite advances in sequence-property prediction, DL models lack interpretability due to several non-linear transformations. This limits their ability to dissect monomers' contribution to the property. Understanding these contributions is essential for identifying critical motifs \cite{tracing3} to design mutations for enhanced protein properties \cite{arakawa20221000, pandey2024sequence}. Thus, there is a great need for a model that elucidates monomer-level contributions while establishing sequence-property relationships.

With the growing emphasis on Explainable AI (XAI) and interpretability \cite{ali2023explainable}, some progress has been made in understanding monomeric contributions in proteins \cite{tracing1,tracing2,tracing3, cam1, cam2, i2,i3} while handling primary sequence as an input. One widely used XAI approach is based on the self-attention mechanism in transformers, where a monomer's contribution is determined by the attention it receives from other monomers in the sequence \cite{wu2020structured}.  Another popular method is Grad-CAM (Class Activation Mapping) \cite{gradcam}, which attributes monomeric contribution to the gradient of the output with respect to the monomer's latent space representation. While attention and gradient-based methods have made some strides in providing interpretability for protein sequences \cite{tracing1,tracing2,tracing3, cam1, cam2}, they come with limitations. The self-attention mechanism has been shown to be unreliable as an XAI tool to identify critical segments in a sequence \cite{attention_not_good_1, attention_not_good_2, attention_not_good_3}. Similarly, Grad-CAM typically relies on the embeddings from certain Transformer or LSTM-based Large Language Models (LLM) which already have layers of non-linear transformations \cite{QuoteTarget, gradcam_graph}. Furthermore, Grad-CAM has been mostly employed for images and graph-based input data \cite{gradcam, gradcam_graph}. Therefore, there exists a gap for a more interpretable and explainable model that can effectively elucidate the contributions of individual monomers within protein sequences. 

Current XAI methods for proteins focus primarily on classification tasks like binding site identification, with limited accuracy (40–45\%) in detecting all the sites \cite{QuoteTarget}. These approaches have not been extended to continuous properties like melting temperature and are mainly used for qualitative analyses, such as identifying critical regions in the vicinity of high-contributing monomers \cite{cam1, cam2, tracing3, QuoteTarget}. Additionally, no comprehensive evaluation strategy exists to validate the monomeric contribution scores in proteins. In contrast, image analysis \cite{vitshap, nazir2023survey, yoshikawa2024explanation, i1} and NLP \cite{jahromi2024sidu} have advanced in quantifying contribution scores using insertion and deletion techniques, which evaluate the impact by systematically adding or removing input segments based on score rankings. To our knowledge, this approach has not been applied to validate monomeric contributions in protein property prediction. We propose leveraging this method to quantify and validate monomeric contribution scores in proteins.

Building on the above discussion, this work aims to develop a more interpretable model for elucidating monomeric contributions within primary sequences and systematically evaluating the contribution scores generated by the model. As the first step, we develop a novel DL model that establishes a primary sequence-property relationship while enabling the tracing of monomeric contributions from predicted outputs. Our analysis involves: (1) benchmarking the predictive performance of our architecture against state-of-the-art (SOTA) models like Transformers, LSTMs, and 1D CNNs; (2) introducing an insertion/deletion-based parameter inspired by image and NLP techniques to evaluate monomeric contribution scores; and (3) using this metric to compare our model's performance with attention- and gradient-based XAI methods. We mainly evaluate our model's performance on diverse datasets, including Anti-Cancer Peptide (ACP) properties \cite{sun2024dctpep}, protein solubility \cite{hon2021soluprot}, binding affinity \cite{gb1}, collagen thermal stability \cite{khare2022collagentransformer}, and Antimicrobial Peptide (AMP) classification \cite{gupta2019feedback}. Our results demonstrate the model's ability to capture monomeric importance across sequences with varying motif sizes and long-range dependencies. Further quantitative analysis reveals that our model exhibits superior explainability for estimating monomeric contribution while maintaining competitive predictive performance. Notably, our model also identifies critical motifs such as RRR, RRI, and RSS that compromise ACP stability, as well as AMP motifs that enhance the likelihood of converting non-AMP sequences to AMP sequences by over 50\% compared to SOTA models. These findings lay a foundation for using our interpretable DL model to design protein-based materials with improved properties.

\section{Materials and Methods}
\subsection{Deep-Learning Framework}
Protein's primary sequence is defined by the sequence of amino acids \cite{branden2012introduction}, which is typically described by a series of letters representing the chemical structure of 20 common amino acids. However, the primary sequence does not contain any explicit information about the secondary and tertiary structure of the protein which defines the shape of the protein in the 3D space \cite{branden2012introduction}. Today, deep learning models such as AlphaFold2/3 \cite{alphafold,alphafold3} enable the prediction of protein structures based on their primary sequences. However, AlphaFold2/3 often exhibits low confidence in predicting amorphous structures and fold-switching regions in proteins \cite{alphafoldfail, alphafoldfail2}. As a result, primary sequence data is more readily accessible as input compared to structural information for proteins, especially for proteins with more disordered regions. Therefore in the literature, there is a lot of work available on predicting protein properties just based on its primary sequence \cite{brandes2022proteinbert, xu2020deep, prottrans, khare2022collagentransformer}. Furthermore, it has been demonstrated that deep learning (DL) models outperform traditional machine learning (ML) methods, such as ridge regression, support vector machines, and random forests, when training datasets exceed 1,000–2,000 samples due to their over-parameterized architectures \cite{sample_efficient}. However, obtaining even 1,000 high-quality training samples remains a challenge in certain specialized applications \cite{arakawa20221000, khare2022collagentransformer, kim2023predicting}. Therefore, there is still a gap for an interpretable sample-efficient deep-learning (DL) model that can estimate the contribution of monomers in the primary sequence. Understanding monomeric contributions will enable the identification of critical motifs within sequences, which are vital for maintaining specific functions or enhancing desired properties. Therefore, we have formulated an interpretable DL model for proteins in this work. 

\begin{figure}[htbp]
  \includegraphics[width=17 cm]{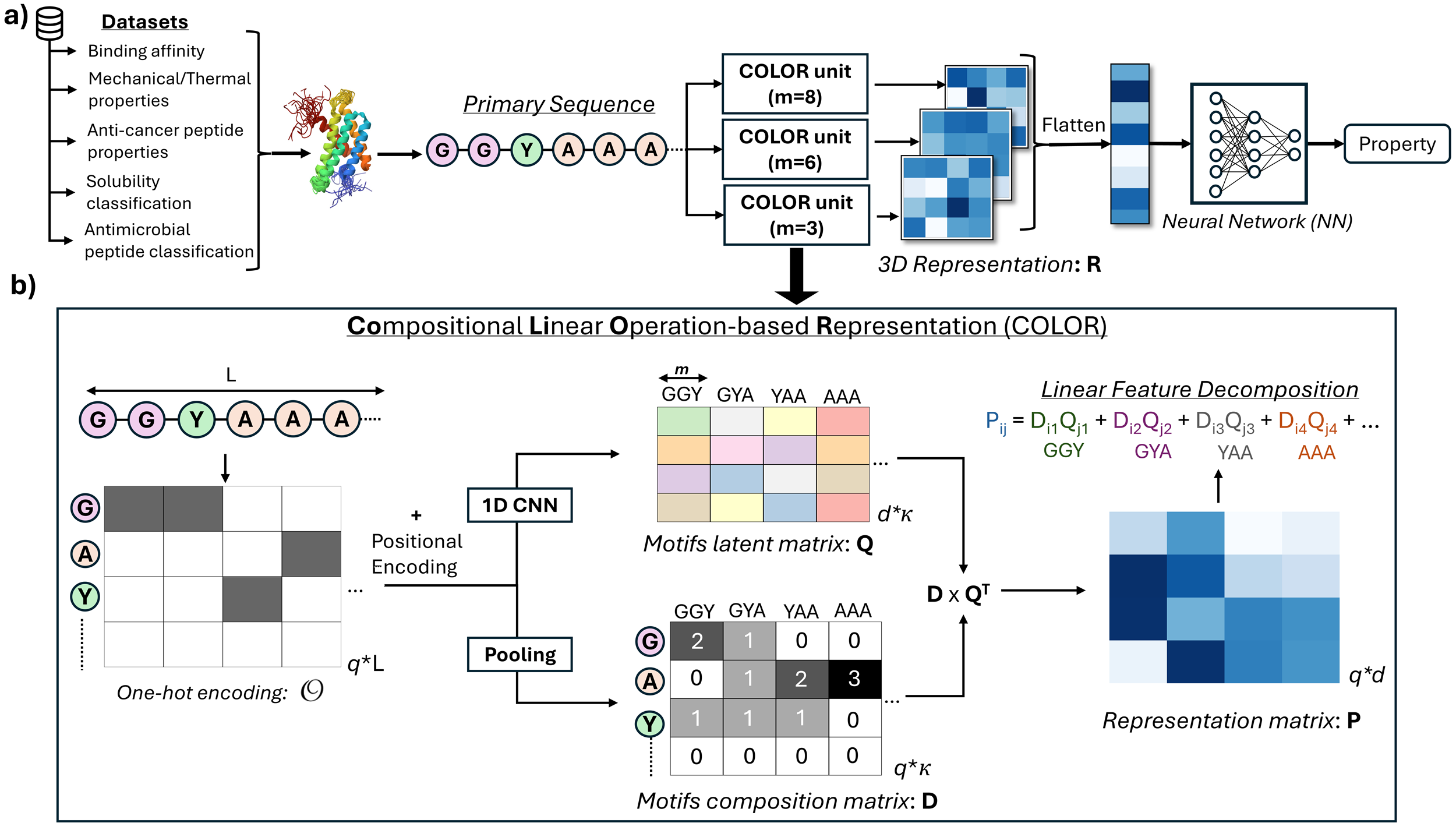}
  \caption{Overall interpretable deep-learning model architecture. (a) The complete data flow for predicting properties from the primary sequence. (b) Detailed components of the COLOR unit, illustrating the linear decomposition of elements in \pp into motifs, thereby showcasing its interpretability. The different \sm values displayed for the COLOR unit in the figure are intended for reference only and can be adjusted based on the user's specific input or application needs.}
  \label{model_arch}
\end{figure}

Our DL model uses a \textbf{Co}mpositional \textbf{L}inear \textbf{O}peration-based \textbf{R}epresentation (COLOR) unit, a key contribution of the work. A COLOR unit consists of 3 modules namely: sequence-to-motif conversion module, motif composition module, and linear weighted summation module. COLOR unit architecture is shown in Fig.\ref{model_arch}. Before describing these modules, we introduce the term \textit{number of qualitative variables} ($\mathrm{q}$), representing the total distinct qualitative variables that can appear in the sequence. For proteins, $\mathrm{q}$ is typically considered to be 21, accounting for the 20 most common amino acids and one additional for uncommon amino acids such as hydroxyproline and hydroxylysine in collagen protein \cite{ricard2011collagen}. Additionally, the term "motif" will be used frequently in the model description and, in this context, refers to any sub-segment of the primary sequence.

\subsubsection{Sequence-to-Motif Module}
This module divides the primary sequence into several motifs using a 1D convolution network (CNN) \cite{kiranyaz20191}. For example, GGYAAA can be divided into motifs such as GGY, GYA, YAA, and AAA of size 3. The motif size ($\mathcal{m}$) can be controlled by regulating the filter size in the 1D CNN \cite{paszke2017automatic}. The filter size controls the number of neighboring amino acids that the 1D CNN considers in front of each monomer for feature extraction. It is important to note that motifs are created by sweeping with a filter of size $\mathcal{m}$ across the primary sequence with the stride of 1. Therefore, the number of motifs ($\kappa$) of size $\mathcal{m}$ obtained from a primary sequence of length $L$ is ($L-\mathcal{m}+1$). 

The 1D CNN uses the one-hot encoded representation of the primary protein sequence as input \cite{ohe}. For example, if the sequence contains the amino acid 'G' at a specific position, its encoding is represented as [1, 0, 0, ...], where the first position in the vector corresponds to 'G'. The one-hot encoding ($\mathcal{O}$ $\in$ $R^{\mathrm{q}*L}$) is the sparse representation of the primary sequence simply capturing the type of qualitative variable at every position in the primary sequence. The 1D CNN divides the primary sequence into $\kappa$ motifs and generates a latent space vector of size $d$ for each motif as follows:
\begin{equation}
    \begin{aligned}
        & \textbf{Q} = f_{1DCNN}(\mathcal{O}),~ \text{where} ~~\textbf{Q} \in \mathbb{R}^{d*\kappa}
    \label{q}
    \end{aligned}
\end{equation}
This step is illustrated in Fig.\ref{model_arch} where each column of the matrix \qq represents a latent vector for a motif.

\subsubsection{Motif Composition Module}
In this part of the COLOR unit, the model captures the composition of each motif i.e., it captures the number of different qualitative variables present in each motif. This is obtained by the pooling operation which takes \oo as input and produces motif composition matrix \dd as follows: 
\begin{equation}
    \begin{aligned}
        & \textbf{D}_{ij} = \sum_{k=j}^{j+m} \mathcal{O}_{ik} 
    \label{pool}
    \end{aligned}
\end{equation}
An example of matrix \dd is shown in Fig.\ref{model_arch}.

\subsubsection{Linear Weighted Summation Module}
Till this stage, the model has converted the primary sequence into motifs and has computed latent space (\qq) and composition (\dd) matrix. However, to predict the property based on the primary sequence it is important to accumulate the impact of all the motifs. Hence in this module of the COLOR unit, a representation matrix \pp is obtained by linearly combining the properties of the motifs in the latent space as follows:
\begin{equation}
    \begin{aligned}
        & \textbf{P} = \textbf{D}\times \textbf{Q}^T
    \label{linear}
    \end{aligned}
\end{equation}
Element $\textbf{P}_{ij}$ captures the linear combination of latent property $i$ (where, i = 1,2...$d$) of motifs weighted by the number of $j^{th}$ (where, j = 1,2...$\mathrm{q}$) qualitative variable in each motif. It is important to note that the size of the matrix \pp is independent of the sequence length $L$, unlike other architectures like Transformers, LSTM, and 1D CNNs, where the output size depends on $L$.

\subsubsection{Positional Encoding}
It is important to note that the linear weighted sum of the property of motifs in the latent space to obtain \pp does not consider the order of occurrence of motifs in the primary sequence. This can lead to the loss of any sequential information and poor prediction. We apply positional encoding (PE) defined as
\begin{equation}
    \begin{aligned}
        & PE_{(l, 2i)} = sin\Bigg(\frac{l}{10000^{2i/\mathcal{q}}}\Bigg) \\ 
        & PE_{(l, 2i+1)} = cos\Bigg(\frac{l}{10000^{2i/\mathcal{q}}}\Bigg)
    \label{positional}
    \end{aligned}
\end{equation}
 to each position $l$ ($l$=1,2...L), and add the resulting PE to \oo before inputting it into the 1D CNN, ensuring the retention of sequential information. In Eq.\ref{positional}, for even \sq, $2i$ assumes on the values \{0,2,4,..,$q$\}, while $2i+1$ takes on the values \{1,3,5,..,$q$-1\}. It should be noted that sine and cosine functions are used for the even and odd positions along the rows of \oo, respectively. We incorporate positional encoding (PE) from Vaswani et al.\cite{vaswani2017attention} into COLOR, avoiding the direct use of the numeric position $l$ as PE. Using $l$ as PE can cause feature scaling issues, particularly for longer sequences ($L$), which can hinder model optimization. Instead, as shown in Eq. \ref{positional}, PE is embedded as a vector of size \sq, allowing the model to effectively capture and retain relative positional information.

\subsubsection{Complete Architecture}
In the above sections, we discussed the method to obtain sequence-length independent representation matrix \pp for a particular motif size $\mathcal{m}$ using the COLOR unit. However, using \pp pertaining to just one motif size,$m$, can be insufficient to fully capture the behavior of the protein since motifs of varying size may contribute strongly to a given property. For this reason, we can use several COLOR units to obtain different \pp based on several motif sizes as shown in Fig.\ref{model_arch}a with the detailed structure of COLOR unit depicted in Fig.\ref{model_arch}b. Different \pp matrix can be assembled into a 3D representation matrix \rr. Subsequently \rr is flattened and fed into a neural network (NN) to make the property prediction. The cardinality of the set $m$, representing the number of COLOR units in \rr, is denoted as $|m|$. It is important to consider that increasing $|m|$ increases the model's trainable parameters, so this should be scaled appropriately with the available training data to prevent overfitting. The process to select $|m|$ and the values of \sm will be discussed later in Sec.\textbf{Hyperparameter Selection}. As a note, we would like to highlight that whenever the term '\textit{COLOR method}' is used in the text, it refers to the DL model based on COLOR units.

\subsubsection{Quantifying Predictability}
To quantify the predictive performance of the proposed deep learning model, we adopt an approach inspired by Bornschein et al. \cite{predictive_metrics}, where the error is plotted as a function of the training dataset size ($N_{T}$). The area under this curve is then utilized as the metric to evaluate the model's predictive capability. Specifically, we consider two distinct areas under the curve to analyze the model's performance across different data regimes. The first metric, $\mathcal{A}$, computed as 
\begin{equation}
    \mathcal{A}  = \int_{0}^{\infty} e(n)dn
\label{pred_metric_a}
\end{equation}
 quantifies the predictive performance across all training data regimes. In the equation, $e(n)$ represents the mean absolute error (MAE) for regression tasks and accuracy for classification tasks. The second metric, $\mathcal{A}_{500}$, computed as 
 \begin{equation}
    \mathcal{A_{500}} = \int_{0}^{500} e(n)dn
\label{pred_metric_b} 
\end{equation}
 focuses on the model's predictive ability in the low-data regime. For regression tasks, lower values of $\mathcal{A}$ and $\mathcal{A}_{500}$ indicate superior model performance. However, for classification tasks, higher values correspond to better model effectiveness.


\subsection{Monomeric Contribution Calculation Method}
Studying the contribution of monomers in the primary sequence can help estimate the segments/motifs responsible for modulating properties within proteins. However, estimating monomeric contribution is not straightforward for DL models due to the added layers of non-linearity applied to the primary sequence during property prediction. The added non-linearity makes it nearly impossible to trace the impact of each position on the property. Thus, there is a need for interpretable DL models to address this challenge effectively, and COLOR offers the interpretability necessary to estimate monomeric contribution. Therefore, in this section, we develop steps to estimate the monomeric contribution scores based on the COLOR unit. We show how the architectural decisions in the COLOR unit make it possible to trace the impact of positions on the property. 

\subsubsection{Estimating Monomeric Contribution using COLOR unit}
Estimating the monomeric contribution based on COLOR discussed in Sec.\textbf{Deep-Learning Framework} is a two-step process. The first step is estimating the importance of all the features in \rr. The second step involves propagating the importance of features in \rr to motifs in matrix \qq. The details of both the steps are as follows: 

\noindent \textbf{Step 1: Feature Importance} \\
After training the model using the architecture shown in Fig.\ref{model_arch}, we first estimate the importance of elements in 3D matrix \rr. However, as discussed in Sec.\textbf{Complete Architecture}, the elements of \rr are the elements of the different 2D matrices \pp. Therefore, for simplicity, let us assume that we are estimating the importance of elements in \pp. For the feature importance study, we use the method called permutation feature algorithm \cite{molnar2020interpretable}. According to this method, to calculate the importance of a feature $\mathrm{P}_{kl}$ in \pp, we do the following:
\begin{itemize}
    \item First calculate the loss (e$_o$) for the test data with the original set of features using the trained DL model. 
    \item Then freeze all the features as in the original set except $\mathrm{P}_{kl}$. Shuffle $\mathrm{P}_{kl}$ among all the test examples. This breaks the relation that the DL model has learned between $\mathrm{P}_{kl}$ and the output. 
    \item Use the trained model to estimate the loss (e$_p$) with the shuffled $\mathrm{P}_{kl}$ feature. 
    \item Importance of $\mathrm{P}_{kl}$ is calculated as $\vert$100*(e$_p$ - e$_o$)/e$_o$)$\vert$ and is indicated by $o_{kl}$
\end{itemize}

The steps outlined above are repeated for all features in \pp. We are only concerned with the relative importance of all features, all $o_{kl}$ values are divided by the maximum importance value across all features. This scales all the $o_{kl}$ between 0 and 1, with 1 indicating the most important feature.

\noindent \textbf{Step 2: Monomeric Contribution Calculation} \\
The importance of elements in \pp does not directly translate into the contribution of motifs in the primary sequence. So, in this step, we formulate the method to propagate the importance of features to monomeric contribution. From Eq.\ref{linear}, every element in \pp can be expanded as 
\begin{equation}
    \begin{aligned}
        & \mathrm{P}_{ij} = \sum_{k=1}^{\kappa} \mathrm{D}_{ik}\mathrm{Q}_{jk}
    \label{pos}
    \end{aligned}
\end{equation}
, where every term $\mathrm{D}_{ik} \mathrm{Q}_{jk}$ in the equation corresponds to one motif, with a total of $\kappa$ motifs. The greater the magnitude of $\mathrm{D}_{ik} \mathrm{Q}_{jk}$, the stronger the influence of the motif on the corresponding $\mathrm{P}_{ij}$. Therefore, based on the magnitude of $\mathrm{D}_{ik} \mathrm{Q}_{jk}$, all the motifs corresponding to $\mathrm{P}_{ij}$ can be ranked from 0 to ($\kappa$-1) with $r_m$ indicating their ranks. Given $\mathrm{P}_{ij}$ and ranks of different motifs, the contribution score $\phi_m$ assigned to each motif is calculated as 
\begin{equation}
    \begin{aligned}
        \phi_m(r_m;\mathrm{P}_{ij}, o_{ij}) = o_{ij}\times(\kappa-r_m)
        \times\frac{|\mathrm{D}_{ik}\mathrm{Q}_{jk}|-\underset{k}{\min}(|\mathrm{D}_{ik}\mathrm{Q}_{jk}|)}
        {\underset{k}{\max}(|\mathrm{D}_{ik}\mathrm{Q}_{jk}|) - \underset{k}{\min}(|\mathrm{D}_{ik}\mathrm{Q}_{jk}|)}
    \label{impa}
    \end{aligned}
\end{equation}
The non-linearity introduced by the neural network (NN) in the model architecture can lead to noisy latent properties ($\mathrm{Q}_{jk}$) for motifs, particularly affecting the $\mathrm{Q}_{jk}$ with smaller magnitudes. Hence, to mitigate the effect of such noise on $\phi_m$, we apply min-max scaling of $\mathrm{D}_{ik} \mathrm{Q}_{jk}$ in Eq.\ref{impa}, effectively reducing the contribution of noisy, smaller $\mathrm{D}_{ik} \mathrm{Q}_{jk}$ values to nearly zero. It is important to note that motifs might repeat themselves within a sequence but are treated uniquely, as the 1D CNN in Eq.~\ref{q} generates distinct latent space representations for each occurrence in \qq. This differentiation is enabled by positional encoding (Eq.\ref{positional}) that is added to \oo to inform the model about the relative position of monomers in the sequence. 

In the method discussed above, handling motif sizes (\sm\!\!) of 5 and 25 is very different. To illustrate, for a primary sequence with length $L$=50, motif sizes ($\mathcal{m}$) of 5 and 25 yield 46 and 26 motifs, respectively.  Consequently, a motif ranked 5$^{th}$ out of 46 should be assigned a higher score than the one ranked 5$^{th}$ out of 26, as the ranking reflects a larger search space for smaller motif sizes. To account for this, we introduce a variable $\lambda$ which indicates the number of positions in the primary sequence that have been assigned a contribution score. Using $\lambda$ we scale the contribution score $\phi_m$ as
\begin{equation}
    \overline{\phi_m} = \phi_m \Bigg(1- \frac{\lambda}{L} \Bigg) 
    \label{impb} 
\end{equation}
Once $\overline{\phi_m}$ is assigned to different motifs, a contribution score will be associated with every position in the primary sequence.

\subsubsection{Quantifying Explainability}
Methods such as Grad-CAM and attention tracing have been used to visualize the important regions in proteins \cite{gradcam, tracing1, tracing2, tracing3, madsen2021evaluating}. However, a systematic evaluation of the monomeric contribution scores ($\overline{\phi}$) within primary sequences remains unexplored in the field of protein property prediction. In contrast, the fields of natural language processing (NLP) and computer vision have established methodologies to assess contribution scores \cite{vitshap, hooker2019benchmark, pham_double_trouble}. In these works, authors mask certain important positions (pixels in the case of an image) and re-train the model with the masked input. The model which is efficient in ranking positions (or pixels) based on contribution score, leads to a higher drop in the performance after masking. Masking a position or pixel refers to zeroing out its contribution to the model’s final output. 

Drawing inspiration from the masking-based method discussed above, we have formulated a similar technique to evaluate the contribution scores. Once the contribution scores $\overline{\phi}$ are calculated, the monomers in the primary sequence are ranked. Subsequently, all monomers are masked except for the top $u$\%, after which the model is re-trained to assess performance. The value of $u$ is incrementally increased, and with each step, the model is re-trained, while the error (or accuracy) on the test data is recorded after each retraining. By plotting the error (or accuracy) as a function of the unmasked percentage ($u$\%), a curve is generated. The area under this curve, $\mathcal{I}$, calculated as  
\begin{equation}
    \begin{aligned}
        & \mathcal{I} = \int_{0}^{100} e(u)du
        \label{inter_metric}
    \end{aligned}
\end{equation}
is used as a quantitative metric to evaluate the explainability of the model. The explainability of the COLOR method will be rigorously evaluated against state-of-the-art XAI models, including Grad-CAM \cite{gradcam}, Attention Tracing \cite{attention_tracing, tracing1, tracing2, tracing3}, and Grad-SAM \cite{gradsam}, with details presented in Sec. \textbf{Related interpretable models for protein sequences} of Supplementary Information. 

\subsection{Dataset}
In the main paper, we present results for 7 unique properties derived from distinct datasets. These datasets include continuous properties analyzed as regression tasks and categorical properties analyzed as classification tasks. The details of all these datasets will follow shortly. We also conduct additional analysis to further demonstrate the robustness of the COLOR method using two toy datasets and a computational silk dataset \cite{kim2023predicting} which are discussed in detail in Sec.\textbf{Supplementary Datasets} in Supplementary Information. 

\subsubsection{Anti-Cancer Peptide (ACP) Properties}
The instability index of the protein captures the intracellular stability of the protein \cite{guruprasad1990correlation}. It can hugely vary based on the primary sequence of proteins. Sun et al. \cite{sun2024dctpep} have constructed a comprehensive dataset documenting the instability index of several ACPs with $L$ ranging between 20-97. In our current work, we utilize this documented instability index for different primary sequences as one of the key datasets. We have also included the Aliphatic \cite{aliphatic} and GRAVY \cite{gravy} index of ACPs documented in the same database \cite{sun2024dctpep} as two different datasets for our current study. The aliphatic index of a protein is defined as the proportion of the volume occupied by aliphatic side chains, which include alanine (A), valine (V), isoleucine (I), and leucine (L). The GRAVY index is the sum of the hydropathy value of all the amino acids, divided by the sequence length ($L$). Including the Aliphatic and GRAVY index in the dataset is crucial for conducting monomeric contribution calculation studies, where the primary sequence is masked based on its contribution score, and the properties are re-evaluated as described in Sec. \textbf{Quantifying Explainability}. In the case of these two datasets, re-evaluating the properties is straightforward due to the availability of an analytical formulation. 

\subsubsection{Collagen Melting Temperature ($T_m$)}
Collagen is one of the most abundant proteins in animals and has numerous applications in the field of medicine \cite{collagen_application}. Due to its ubiquitous applications, collagen needs to be thermally stable. Yu et al.\cite{khare2022collagentransformer} have experimentally gathered the melting temperature of 633 different primary sequences of collagen to investigate their thermal stability. The higher the melting temperature, the greater the stability of collagen. 

\subsubsection{GB1 binding affinity}
Olson et al. \cite{gb1} developed a dataset containing an experimentally calculated binding affinity of double mutated protein G domain B1 (GB1) to immunoglobulin G fragment crystallizable (IgGFC). This dataset contains the binding affinity of mutated sequences of GB1 protein. 

\subsubsection{Soluprot}
Protein solubility is crucial for the production of various therapeutics \cite{hon2021soluprot}, making it an essential property to predict. Hon et al. \cite{hon2021soluprot} used TargetTrack \cite{berman2017protein} to extract the data on the solubility of proteins in E.coli. The dataset consists of 11,436 training data and 3,100 test data.  

\subsubsection{Antimicrobial Peptide (AMP) Classification}
AMPs are small molecular peptide that possesses anti-microbial functions against a broad range of microorganisms such as bacteria, fungi, parasites, and viruses. Gupta et al. \cite{gupta2019feedback} have curated a dataset of 5200 short peptides, with 2,600 experimentally verified as AMPs, while the remaining sequences are non-AMPs.


The data split for all the datasets mentioned above is given in Tab.\ref{datasplit}.

\begin{table}[H]
\begin{tabular}{|l|c|c|c|}
\hline
\multicolumn{1}{|c|}{Dataset}                                                & Training & Validation & Test \\ \hline
\begin{tabular}[c]{@{}l@{}}ACP Aliphatic, GRAVY, Instability \\\end{tabular} & 850      & 150        & 150  \\ \hline
\begin{tabular}[c]{@{}l@{}}Collagen Melting Temperature\end{tabular}       & 506      & 63         & 64   \\ \hline
\begin{tabular}[c]{@{}l@{}}GB1 Binding Affinity\end{tabular}              & 10000    & 5255       & 5308 \\ \hline
Soluprot                                                                     & 8336     & 3100       & 3100 \\ \hline
\begin{tabular}[c]{@{}l@{}}AMP Classification\end{tabular}                 & 3200     & 1000       & 1000 \\ \hline
\end{tabular}
\caption{Data split for different datasets used in the current study.}
\label{datasplit}
\end{table}

\subsection{Hyperparameter Selection}
 
The current model architecture enables tuning of various hyperparameters, allowing control over the model size and its capability to capture protein behavior, as assessed by metrics $\mathcal{A}$ and $\mathcal{A_{500}}$. The size $d$ of the latent space vectors in \qq can be tuned to be large enough to capture motif properties in the latent space, but not so large as to increase the number of trainable parameters unnecessarily. In the current work, $d$ values ranging from 4-20 are selected based on the observed variation of $\mathcal{A}$ across different $d$ values. A parametric analysis of $d$ is presented in Sec. \textbf{Parametric Study}. The analysis reveals that $d$ is influenced by \sq, with higher \sq generally requiring larger $d$ values. Notably, across the datasets mentioned, the model handles \sq values ranging from 4 to 20, as detailed in Supplementary Table 1. One of the important variables for obtaining matrix \qq is \sm which controls the size of the motifs as discussed in Sec.\textbf{Sequence-to-Motif module}. The value of \sm should be selected based on the application. For example, Glycine (G) appears at every third position in collagen protein, suggesting motifs of size 3 \cite{colgen}; hence choosing \sm to be 3 is a logical choice. In many cases, the exact value of \sm or $|m|$ suitable for a specific application may be unknown. However, a sensitivity analysis can be performed to identify \sm and $|m|$ values that achieve an optimal balance between predictability and explainability, as detailed in Sec.\textbf{Parametric Study}. Based on the above discussion, several factors influence the model's performance. We provide key guidelines and recommendations in Sec.\textbf{ Strategies for Enhancing Model Performance} of Supplemental information to aid in optimizing the architecture.


\section{Results and Discussion}

\subsection{Predictive Capability}

\begin{figure}[htbp]
  \includegraphics[width=16 cm]{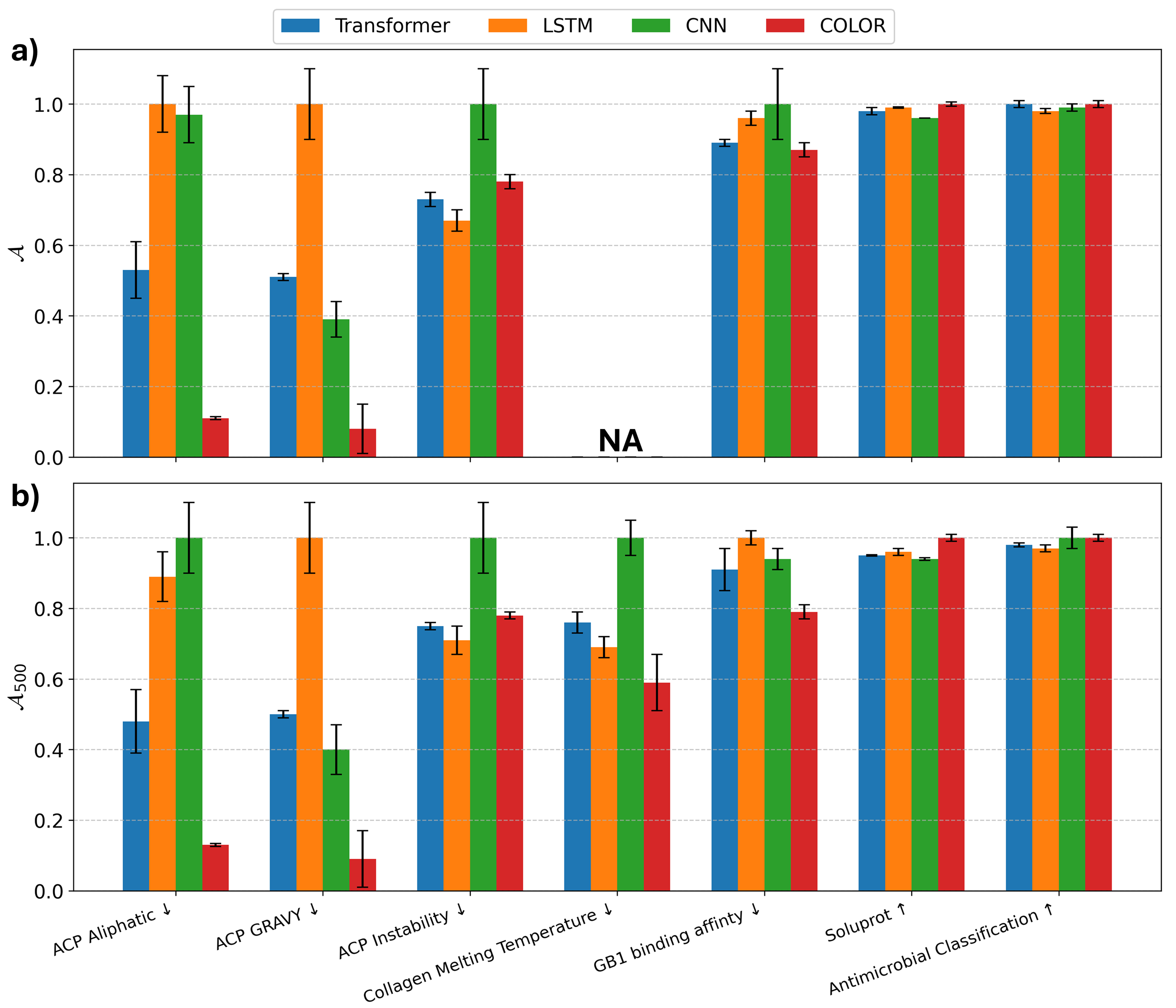}
  \caption{Comparison of the predictive capability of different supervised models. Figure a) shows the comparison of $\mathcal{A}$ and b) illustrates the comparison of $\mathcal{A_{500}}$ obtained for different datasets. The arrows $\uparrow$ and $\downarrow$ in front of dataset names indicate whether higher or lower values are better, respectively. The results in the table are \textit{normalized} using the highest values of the corresponding dataset.}
  \label{predictive}
\end{figure}

To study the predictive capability of the model proposed in Sec.\textbf{Deep-Learning Framework}, we test it on all the datasets discussed in Sec.\textbf{Dataset} and compare the results with the current state-of-the-art (SOTA) models such as Transformers, LSTM, and 1D CNN. We also present additional analysis on two toy datasets and a silk dataset in Sec.\textbf{Comprehensive Predictive Capability} in Supplementary Information. The details of the models used for various datasets are given in Supplementary Tables 1 and 2. We employ the area under the curve metrics, namely $\mathcal{A}$ and $\mathcal{A_{500}}$, as introduced in Sec.\textbf{Quantifying Predictability}, to compare the predictive capabilities of various models. The $\mathcal{A_{500}}$ metric is specifically used to evaluate the model's predictive capability under data-constraint scenarios (smaller training data). The metrics $\mathcal{A}$ and $\mathcal{A_{500}}$ are derived from the curve of MAE (for regression) or Accuracy (for classification) plotted against the number of training samples. These curves are shown in Supplementary Figure 4. A comparison of models is shown in Fig.\ref{predictive}. COLOR outperforms the next best SOTA model by 1-79\% across various datasets except for the ACP Instability dataset, where it performs worse by 16\%. The lower performance in the case of the instability dataset can be attributed to the lower degree of non-linearity in COLOR as it aggregates the contribution of various motifs through a simple linear operation as shown in Eq.\ref{linear}. However, it will be demonstrated in a later section, that despite the lower predictive performance in the case of the two datasets, the model excels at capturing the monomeric contribution within the primary sequence; emphasizing the higher explainability offered by COLOR. 

\subsection{How Explainable is the model?}

Having shown that our current model has a good predictive capability in the above section, we proceed to show its capability to capture monomeric contribution in the primary sequence as discussed in Sec.\textbf{Monomeric Contribution Calculation Method}.  The multiple layers of non-linearity in DL models render them non-interpretable, making it difficult to trace the contribution of each position in the primary sequence to the predicted property. Therefore, it is safe to say that for the model to predict the monomeric contribution, it needs to be interpretable. To that end, we use our method discussed in Sec.\textbf{Monomeric Contribution Calculation Method} to predict the monomeric contribution and quantify it using the metric $\mathcal{I}$ introduced in Eq.\ref{inter_metric}. The calculation of $\mathcal{I}$ involves re-training the model after unmasking $x\%$ of the monomers in the sequence. However, re-training is not required for ACP GRAVY and Aliphatic index datasets, as their output $y$ can be computed analytically. For this study, we dropped the GB1 binding affinity and Soluprot dataset. We do not consider the GB1 binding affinity dataset for this study as it contains highly similar sequences with only two mutations in the wild-type protein. Additionally, we also drop the Soluprot dataset as it contains noisy labels for the solubility of proteins \cite{hon2021soluprot} leading to lower model accuracy as shown in Supplementary Figure 4.

In this study, we use random assignment of the monomeric contribution scores within the primary sequence as one of the baseline methods. This random method provides a baseline against which our method should perform better, indicating that it has learned some meaningful information about the sequence. A comparison of COLOR with other SOTA models along with random baseline is given in Fig.\ref{interpretability}. The $\mathcal{I}$ values reported are obtained from the curves shown in Supplementary Figure 5. All results are normalized using the results from the random method. The figure shows that the COLOR method achieves the highest performance, outperforming the next-best method by 1–38\% across datasets, with an average gain of 22\%. Notably, our approach consistently outperforms random baselines, a result not guaranteed by other SOTA models. The above observations suggest that our method offers more interpretability, making it more effective in estimating the monomeric contribution within the primary sequence. Additional analyses highlighting the explainability of the COLOR are presented in Sec.\textbf{Comprehensive Explainability} in Supplementary Information. 

\begin{figure}[htbp]
  \includegraphics[width=16 cm]{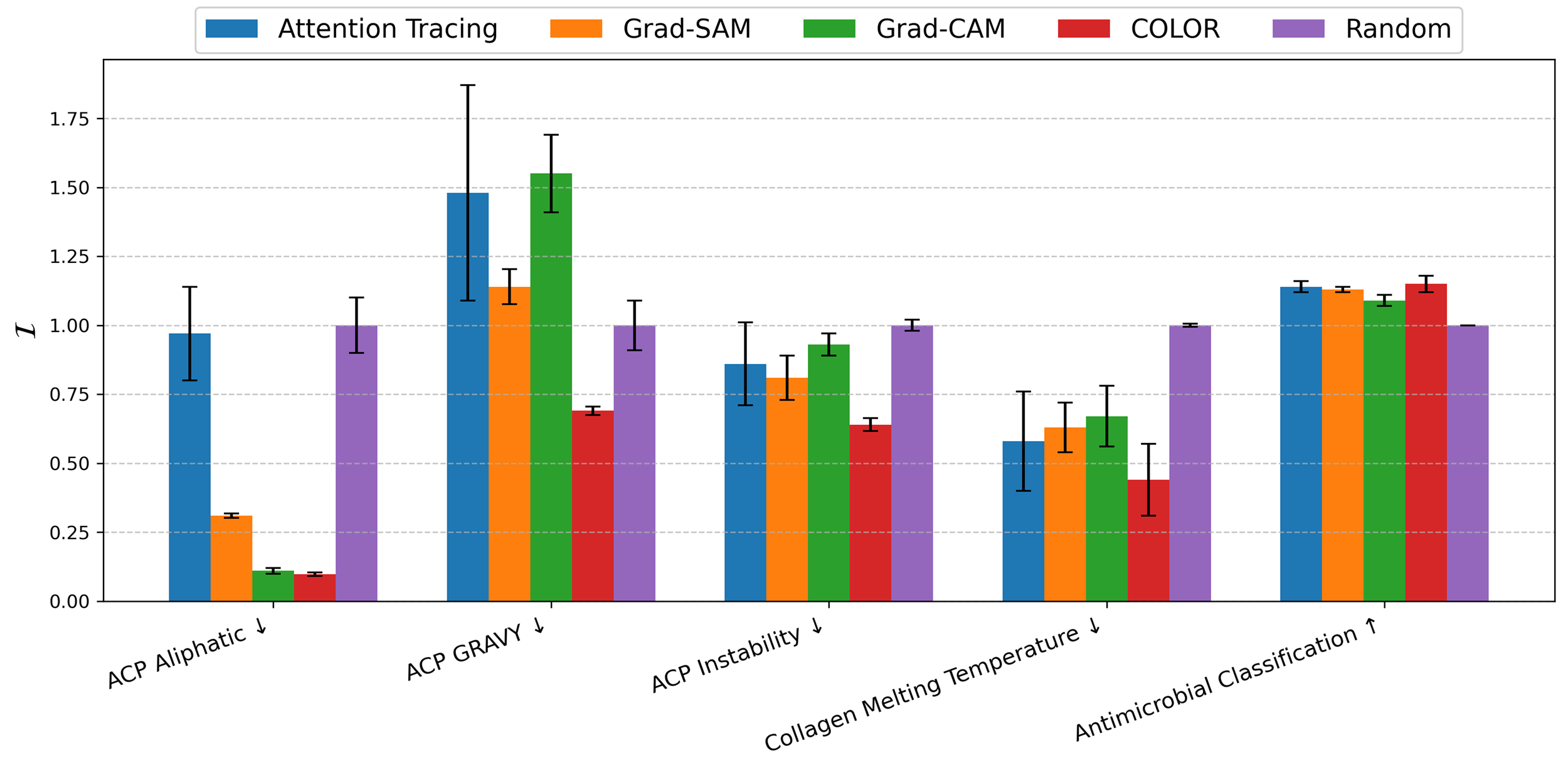}
  \caption{Comparison of explainability offered by different XAI models. The results are normalized using the results from the \textit{Random method} of the corresponding dataset. The arrows $\uparrow$ and $\downarrow$ in front of dataset names indicate whether higher or lower values are better, respectively.}
  \label{interpretability}
\end{figure}

\begin{figure}[htbp]
  \includegraphics[width=16 cm]{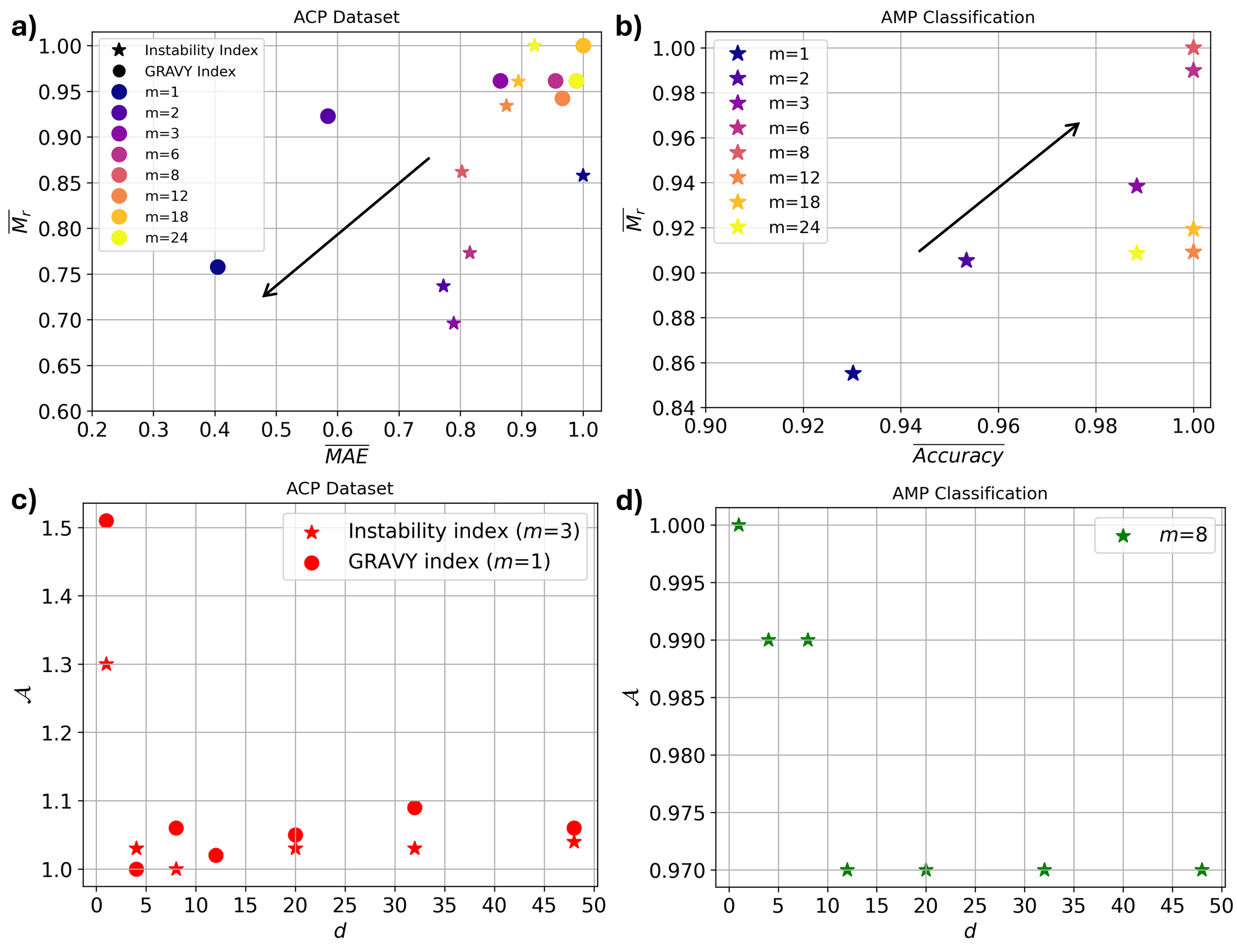}
  \caption{Parametric study depicting the effect of $m$ and $d$ on COLOR's predictive performance and explainability. Figures (a) and (b) show the effect of $m$ on the ACP dataset (Instability and GRAVY index) and AMP classification dataset, respectively, with mean results from three independent runs. Black arrows indicate the direction in which optimal values should trend. Figures (c) and (d) depict the effect of $d$ on predictive performance, where red and green markers indicate lower and higher values corresponding to better model performance, respectively.}
  \label{para_m_d}
\end{figure}

\subsection{Parametric Study}

As discussed in Sec.\textbf{Complete Architecture}, the motif size \sm and the dimensionality of the latent space representation $d$ for each motif are key parameters that define a COLOR unit. Hence, in this section, we conduct a parametric study to examine the impact of varying $m$ and $d$ on the model's predictive performance and explainability. To quantify explainability in the parametric study, we introduce a scaled parameter $\overline{M_{r}}$, defined as 
\begin{equation}
    \begin{aligned}
     M_r = & \text{Predictive performance with only 20\% sequence unmasked} \\
     & \overline{M_r} = \frac{M_r}{\text{max}_{m\in \mathcal{S_m}} (M_r)}, ~\text{where,} ~\mathcal{S_m}=\text{[1,2,3,6,8,12,18,24]}
     \end{aligned}
    \label{mr}
\end{equation}
While computing $M_r$ in Eq.\ref{mr}, the monomers within the top 20\% based on their contribution scores are unmasked, while the remainder of the sequence is masked. To evaluate predictive performance, we utilize scaled MAE ($\overline{MAE}$) for regression tasks and scaled accuracy ($\overline{Accuracy}$) for classification tasks. The scaling process for MAE and accuracy follows the same approach as outlined in Eq.\ref{mr}. Furthermore, to examine the effect of $m$, we fix $|m|$ to 1 and vary the value of $m$. This approach enables us to isolate the influence of motif size on the performance of COLOR.  

In Fig.\ref{para_m_d}a\&b, we show the effect of \sm on the model's predictive performance and explainability. For the parametric study in regression tasks, we selected two properties (Instability index and GRAVY index) from the ACP dataset. The GRAVY index was specifically chosen because $m=1$ is sufficient for accurate prediction, making it an interesting case to explore the impact of increasing $m$ on predictive performance. It is evident from Fig.\ref{para_m_d}a\&b that explainability drops while using \sm$\geq$12 in the case of instability index and AMP classification. This can be attributed to the fact that in the COLOR method, the contribution score $\overline{\phi}$ is assigned at the motif level as per Eq.\ref{impa} \& \ref{impb}. This means that while working with larger \sm values, the model can end up assigning a higher contribution score to a larger motif of which only a smaller segment is important and the rest of the motif is insignificant.
For the same reason,  in the case of the GRAVY Index, which is independent of any interactions between neighboring monomers in the sequence, using any $m>$1 adversely affects explainability as evident from Fig.\ref{para_m_d}a. It can also be noted that in the case of instability index and AMP classification, the model's predictive capability is lower while using \sm=1. This is because the model does not consider any neighboring monomers while generating matrix \qq and linearly combining the effect of various monomers in Eq.\ref{linear} might not be sufficient to capture the effect of neighboring monomers in the sequence. 

There is also a subtle but important difference in the effect of $m$ on ACP instability and AMP classification dataset. For the ACP instability dataset, a smaller motif size (3$<m<$6) yields optimal performance, whereas the AMP classification dataset requires a larger motif size ($m$=6 or 8) for better results. This difference can be attributed to the nature of the datasets: the AMP dataset consists of sequences made up of nucleotide bases (A, G, C, and T), where every three nucleotides correspond to a single amino acid. Since amino acids are crucial for protein function, a larger motif size in nucleotide sequence is necessary to capture 2–3 amino acids in each motif, thereby improving both predictive accuracy and explainability.

For a fixed value of \sm, varying $d$ can lead to different predictive capabilities. Since $d$ determines the size of the vector representing each motif in matrix \qq, adjusting it primarily impacts the number of tunable parameters, thus influencing the model's predictive capability. To study the impact of $d$, we first fix \sm to be 1, 3, and 8 for the GRAVY index, Instability index, and AMP classification datasets respectively based on the results shown in Fig.\ref{para_m_d} a\&b. In Fig.\ref{para_m_d} c\&d, we show the variation of $\mathcal{A}$ (see Eq.\ref{pred_metric_a}) for ACP and AMP datasets. From the figure, it can be noted that in the case of the ACP dataset, choosing $d>$4 is a robust choice for better predictive performance. On the other hand, for the AMP dataset, $\mathcal{A}$ for all $d$ values are within 3\% of each other. The better performance observed with lower $d$ values in the AMP dataset could be attributed to the composition of the sequences, which consist of nucleotides (\sq = 4), necessitating a lower dimensionality ($d$) for effective representation. This study suggests that users may consider setting $d$ proportional to \sq. 
   



\subsection{Interpretability through the lens of human intuition}

\subsubsection{Is latent space representation meaningful?}

To study whether COLOR learns a meaningful latent representation of motifs in matrix \qq, we designed two tasks: in the first, we trained the model to predict the sum of monomeric hydropathy \cite{hydropahthy}; in the second, we trained it to predict the sum of the isoelectric (pI) point \cite{pi}. The pI point of a monomer is reflective of its charge. As the hyperparameters, we fix \sm and $d$ to be 1, indicating that we generate the latent representation for every single monomer rather than motifs. We also replace the neural network shown in Fig.\ref{model_arch} by a simple sum of all the elements of \pp (i.e., y$^p$=$\sum_{i,j} \textbf{P}_{ij}$). After training, we expect the latent space representation learned for each monomer to converge to its respective hydropathy (in the first task) and pI values (in the second task). When comparing the actual monomeric hydropathy and pI values with their corresponding latent space representations from \qq, we achieve an $R^2$ of 1.0, indicating that the model learns meaningful and application-specific representations.

\subsubsection{Does the contribution score reflect expected patterns?}

\begin{figure}[htbp]
  \includegraphics[width=16 cm]{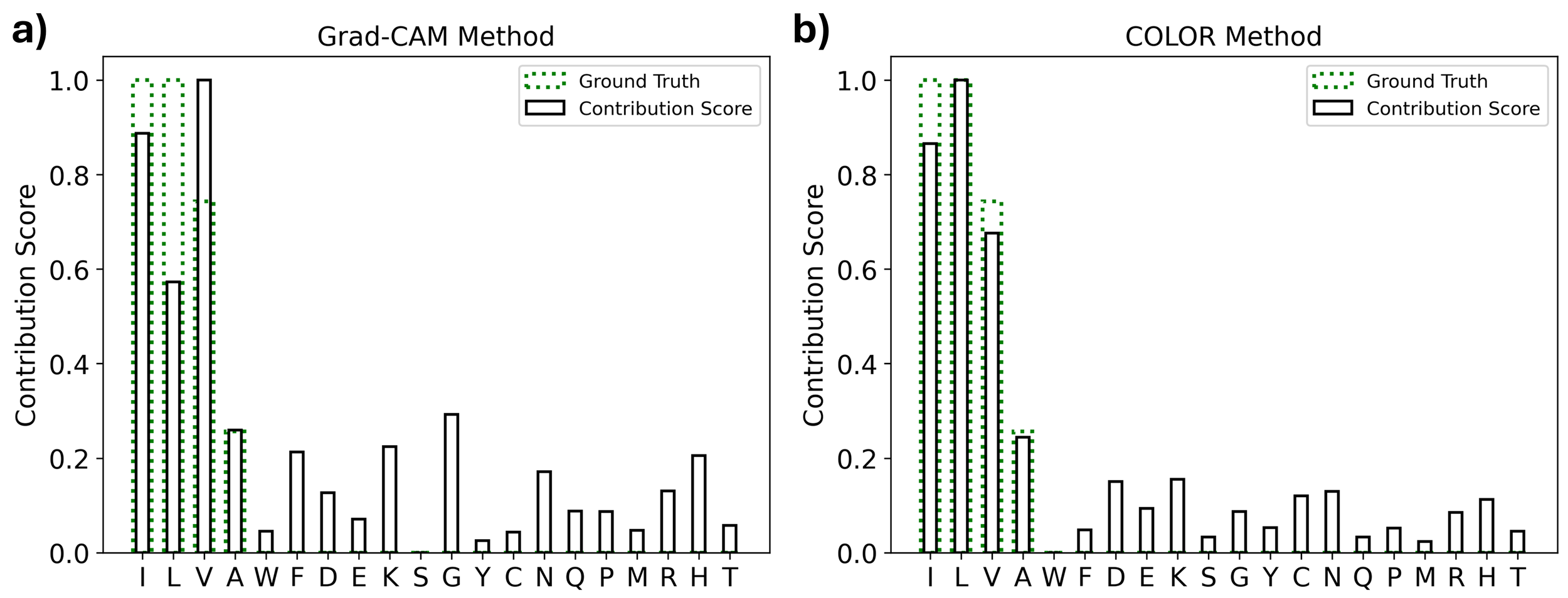}
  \caption{Contribution score assigned to every amino acid when predicting the aliphatic index. a) Contribution score assigned using the Grad-SAM method, and b) using COLOR.}
  \label{aliphatic}
\end{figure}

In this section, we study if the contribution score assigned to different amino acids is proportional to their actual contribution to the output $y$. First, we consider the example of ACP aliphatic index prediction. Given a primary sequence, the aliphatic index can be analytically obtained as
\begin{equation}
    \begin{aligned}
    \text{Aliphatic Index} = \chi(A) + 2.9\chi(V) + 3.9 (\chi(I) + \chi(L))
    \end{aligned}
    \label{aliphatic_eqn}
\end{equation}
, where $\chi$ is the amino acid compositional fraction. According to this equation, amino acids isoleucine (I) and leucine (L) have the highest significance, followed by valine (V) and alanine (A). The mean contribution scores COLOR assigns to every amino acid closely follow the expected trend, as shown in Fig.\ref{aliphatic}b. It should be noted that there are non-zero contribution values for amino acids that are expected to have zero contribution. This arises from the noisy latent vectors learned due to the non-linearity introduced by the neural network in Fig.\ref{model_arch}. In contrast to the COLOR method, the contribution scores from the Grad-SAM method (refer Fig.\ref{aliphatic}a), although the second-best interpretable approach for this dataset as indicated in Fig.\ref{interpretability}, exhibit notable deviations from the expected trend, particularly underestimating the contribution of amino acid A. It is noteworthy that both COLOR and the transformer-based model exhibit excellent predictive capabilities, achieving $R^2>$0.99 in both instances. Consequently, the variation in the contribution scores presented in Fig. \ref{aliphatic} can be attributed solely to the explainability of the respective models.

\begin{figure}[htbp]
  \includegraphics[width=16 cm]{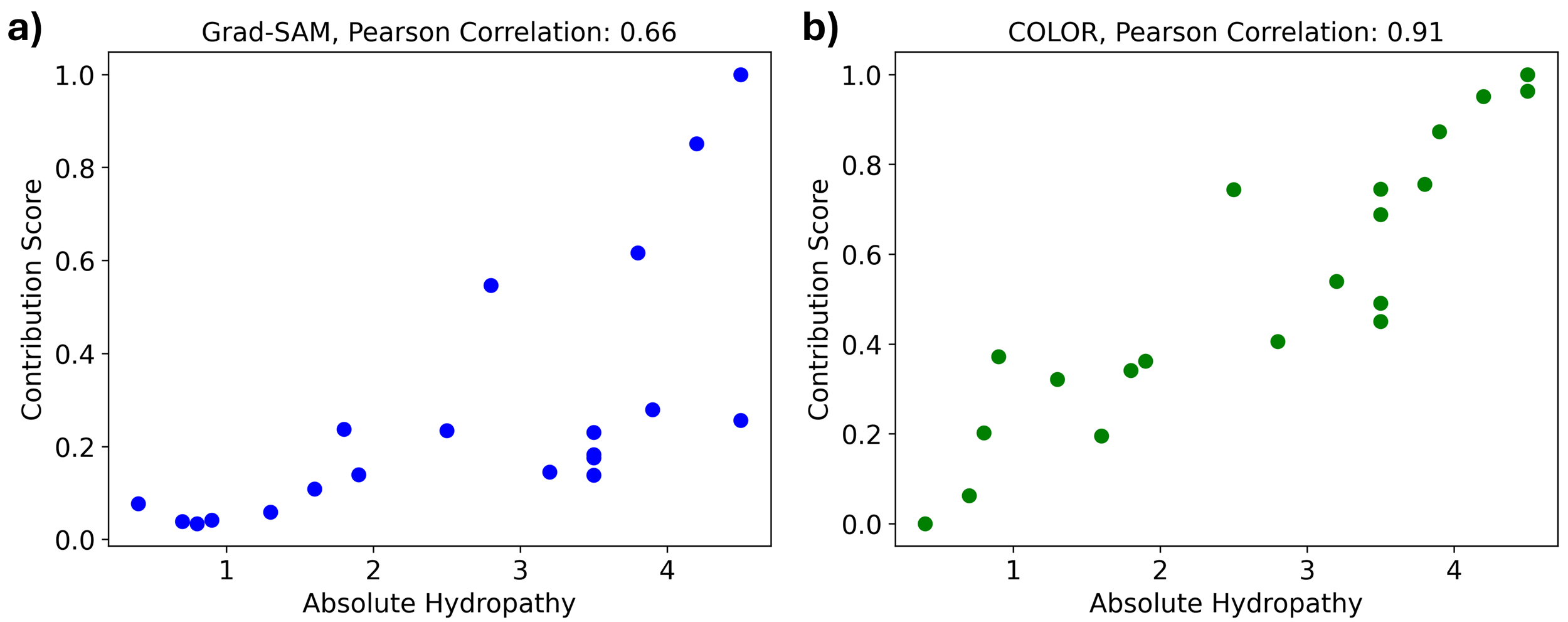}
  \caption{Comparison of contribution scores as a function of monomer hydropathy. a) contribution scores obtained using the Grad-SAM method, and b) contribution scores obtained using COLOR method.}
  \label{gravy}
\end{figure}

We perform a very similar study with the ACP GRAVY index prediction dataset. Analytically, the GRAVY index is the summation of the hydropathy of all monomers in the primary sequence. Fig.\ref{gravy} shows the comparison between the contribution scores and absolute hydropathy of amino acids, wherein we anticipate a strong correlation between the two variables. In Fig.\ref{gravy}a \& b, the contribution scores are obtained using Grad-SAM (second best interpretable model) and our method respectively. The contribution scores from COLOR have $\sim$35\% higher correlation with the absolute hydropathy value, highlighting the effectiveness of our approach in accurately capturing the significance of various amino acids. It is again important to note that both COLOR and the transformer-based model exhibit high predictive capabilities, with  $R^2>$0.99. Therefore, the differences illustrated in Fig. \ref{gravy} arise from the varying explainability of the two models.

\subsection{Application in motif identification}
\begin{figure}[htbp]
  \includegraphics[width=16 cm]{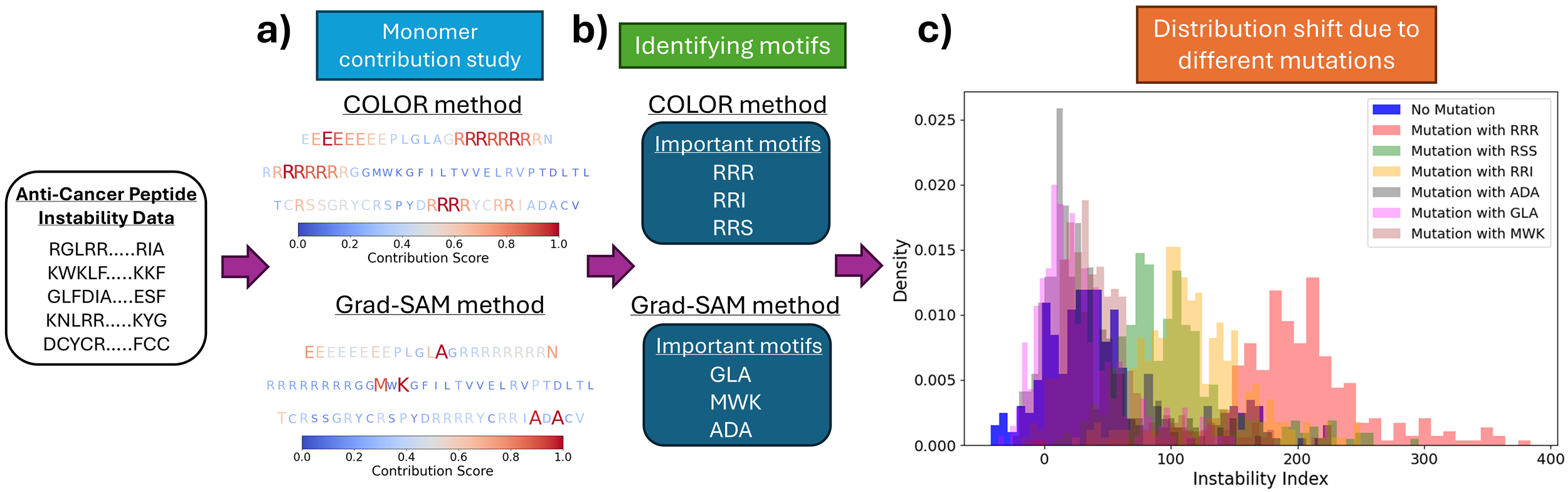}
  \caption{Motif identification study for Anti-Cancer Peptide (ACP) instability index data. a) The contribution of each monomer in the three most unstable peptides (highest instability index) from the test dataset, calculated using the COLOR and Grad-SAM methods. b) Motifs ($i_{un}$) of size 3 are down-selected based on the contribution scores from both methods. c) Illustrates the distribution shift in the instability index of the mutated ACP sequences $\tilde{x}_{t}$. The distribution shift is more pronounced when using motifs RRR, RRI, and RSS for mutation, indicating that the COLOR method has identified more impactful motifs.
}
  \label{acp_motif}
\end{figure}

Having quantitatively demonstrated the explainability of COLOR, we now extend its application to motif identification. Given the rise of anti-cancer peptides for cancer therapies, in this section, we identify the motifs responsible for its instability. We first choose the three most unstable peptides from the ACP Instability index dataset and study the monomeric contribution using Grad-SAM and COLOR as shown in Fig.\ref{acp_motif}a. Grad-SAM was chosen as the method for comparison as it is the second-best interpretable method for instability dataset as shown in Fig.\ref{interpretability}. Subsequently, based on the contribution score, we identify the three most important motifs ($i_{u}$) in the case of both the methods as shown in Fig.\ref{acp_motif}b. To validate the impact of identified motifs, all the test sequences, $x_{t}$, are mutated to $\tilde{x}_{t}$ at the three most important positions ($p_m$) using $i_{un}$. In short, $\tilde{x}_{t}$ = $r$($x_{t}$, $i_{u}$, $p_m$), where $r$ represents the mutation of $x_{test}$ at positions $p_m$ using motif $i_{u}$. For a fair comparison between the two methods, we conduct multiple scenarios: in half of the scenarios, the mutation positions are determined based on contribution scores from the Grad-SAM method, while in the other half, they are selected using the COLOR method. To further avoid any bias, the instability index of $\tilde{x}_{t}$ is calculated using both methods. The distribution of the instability index of $\tilde{x}_{t}$ is shown in Fig.\ref{acp_motif}c in comparison with the distribution of the instability index of $x_{t}$. The shift is dominant when mutated with $i_{u}$ identified using the COLOR method, reinforcing its capability to identify impactful motifs. This study also demonstrates that motifs RRR, RSS, and RRI significantly compromise the stability of ACP.

\begin{figure}[htbp]
  \includegraphics[width=16 cm]{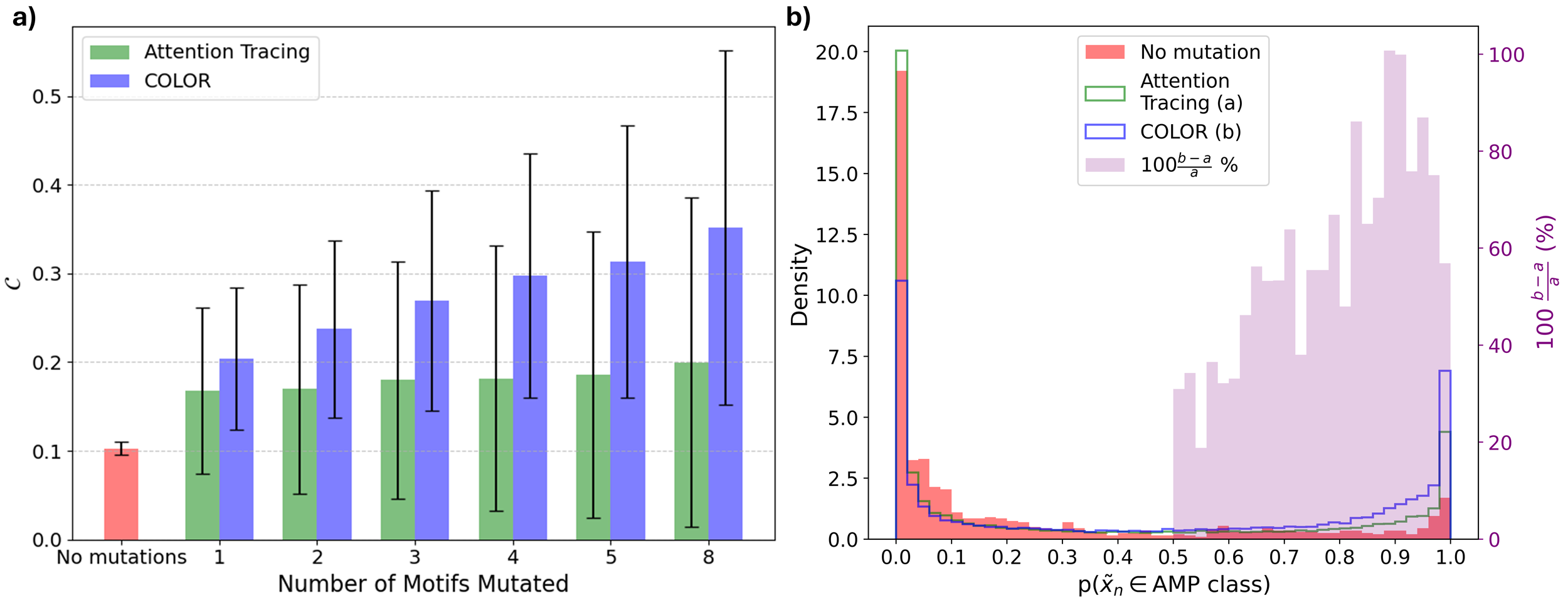}
  \caption{Motif identification and mutation study on Antimicrobial Peptide (AMP). a) Depicts the variation of $\mathcal{C}$ with respect to the number of mutations ($|p_m|$) introduced in $x_{n}$. b) Depicts the distribution of mutated sequences p($\tilde{x}_{n}\in$AMP class) obtained from Attention tracing and COLOR in comparison with the distribution of non-mutated sequences p($x_{n}\in$AMP class). It also shows the \% improvement that the COLOR method offers over Attention tracing in the region p$>$0.5. 
}
  \label{amp_motif}
\end{figure}

As highlighted in Sec.\textbf{Dataset}, AMPs possess broad-spectrum antimicrobial capabilities making them promising for treating drug-resistant microorganisms \cite{papagianni2003ribosomally}. Previous studies have utilized generative ML models to design AMPs with better physicochemical properties \cite{gupta2019feedback, das2021accelerated}, underscoring the significance of understanding and designing newer AMPs. Therefore, in this study, we aim to identify key motifs in AMP sequences and validate their impact through mutation analyses. We first identify key motifs ($i_{a}$) in AMP sequences using Attention tracing (second-best explainable model, see Fig.\ref{interpretability}) and COLOR. To evaluate the impact of $i_{a}$ identified by the two methods, we first select specific positions ($p_m$) for mutation in non-AMP sequences. Similar to the study with ACPs, we conduct multiple scenarios: in half of the scenarios, the mutation positions are determined based on contribution scores from the attention-tracing method, while in the other half, they are selected using the COLOR method. Subsequently, all the non-AMP sequences in the test data, $x_{n}$, are mutated to $\tilde{x}_{n}$, where $\tilde{x}_{n}$=$r$($x_{n}$, $i_{a}$, $p_m$). As in the mutation study on ACPs, $r$ represents the mutation of position $p_m$ with motif $i_{a}$ in $x_{n}$. It is important to note that for a given $i_{a}$, $p_m$ can also be a list for facilitating mutations at multiple positions.  Additionally, since $\tilde{x}_{n}$ is derived from $x_{n}$, their cardinality remain the same for a given $i_{a}$ and $p_m$, i.e., $|x_n|$=$|\tilde{x}_{n}|$. Following the mutations, the probability of a $\tilde{x}_{n}$ belonging to the AMP class, p($\tilde{x}_{n}\in$AMP class), is recalculated using both models to eliminate potential bias toward either method. To quantify the impact of mutation(s), we introduce the variable $\mathcal{C}$ defined as
\begin{equation}
    \begin{aligned}
    \mathcal{C} = \frac{|\{\tilde{x}_{n} | p(\tilde{x}_{n}\in\text{AMP class})>0.8  \}|}{ |\tilde{x}_{n} | } 
    \end{aligned}
    \label{xnon}
\end{equation}
The term $\mathcal{C}$ represents the fraction of non-AMP sequences that exhibit a high probability ($>$0.8) of being classified as AMPs following mutation(s).

Fig.\ref{amp_motif}a shows the variation of $\mathcal{C}$ with respect to the number of mutations introduced in the non-AMP sequences. The higher $\mathcal{C}$ values observed for motifs identified from AMP sequences using the COLOR method reflect its effectiveness in identifying critical motifs in the sequences. Overall, based on $\mathcal{C}$ values, COLOR demonstrates a 53\% mean improvement in the likelihood of converting a non-AMP sequence into an AMP sequence compared to the attention tracing method. Fig.\ref{amp_motif}b depicts the overall distribution of p($\tilde{x}_{n}\in$AMP class) calculated for mutated non-AMP sequences after 8 mutations using both the methods, compared with the distribution without any mutations (p($x_{n}\in$AMP class)). Both methods increase the density in the 0.8-1.0 range; however, the COLOR method demonstrates a higher impact. COLOR method also leads to a substantial decrease in the density in 0-0.1 region. Conversely, the Attention tracing method increases the density in 0-0.1 region, highlighting that certain motifs ($i_{a}$) identified by the method negatively impact the likelihood of converting a non-AMP sequence into an AMP sequence. Sequences with p($\tilde{x}_{n}\in$AMP class)$>$0.5 are classified as AMPs. Therefore, we also present the improvement offered by the motifs identified using the COLOR method over the attention tracing method in the p($\tilde{x}_{n}\in$AMP class)$>$0.5 range. The improvement plot again highlights mean 50\% higher likelihood of converting a non-AMP sequence into an AMP when motifs identified by the COLOR method are used for mutations. In Sec.\textbf{Motif Identification in a Toy Dataset} of the Supplementary Information, we further demonstrate COLOR's enhanced ability to accurately identify motifs in a toy dataset with known critical motifs.

\subsection{Conclusion}

The primary sequence of a protein, which predominantly determines its functions, is more readily accessible than structural information. Consequently, significant efforts have been devoted to establishing primary sequence-property relationships. Deep learning models have proven to be very powerful as the surrogate to establish primary sequence-property relationships in proteins. However, these models contain layers of non-linear transformation, making them uninterpretable. Lack of interpretability makes it difficult to estimate the contribution of each monomer in the sequence which is crucial for the insights about the critical regions in proteins. With the recent efforts towards Explainable AI (XAI), attention and gradient-based methods have been developed to dissect monomeric contribution in proteins. However, current XAI methods primarily focus on classification tasks, such as identifying binding sites in sequences, with limited exploration of continuous properties like melting temperature or mechanical properties. Additionally, these XAI methods have also been shown to have limitations in correctly identifying the important parts of the sequential input in the literature. 

To address the gap for an interpretable model to estimate monomeric contribution, we, rather than improving the existing methods, developed a novel DL model named COLOR in which every step is interpretable to trace back the monomeric contribution from the predicted output. Taking inspiration from the field of NLP and image processing, we also formulated a metric $\mathcal{I}$ to measure the effectiveness of the model in capturing the monomeric contribution. Firstly, we evaluate the predictive capabilities of COLOR against SOTA models, including transformers, LSTM, and 1D CNN. COLOR demonstrates competitive performance, specifically outperforming these SOTA models on 7 out of 10 datasets in the low training data regime ($N_{T}$$<$500), underscoring its superior data efficiency. Subsequently, the rigorous quantitative analysis revealed that COLOR also achieves 22\% higher explainability than the other XAI methods applied to protein data. Notably, COLOR achieves enhanced explainability without compromising the predictive capability.

We further extend the capability of COLOR towards identifying critical motifs in the primary sequence. By analyzing the monomeric contribution scores of anti-cancer peptide (ACP) sequences, we demonstrate that COLOR effectively identifies key motifs—RRR, RRI, and RSS—that significantly compromise ACP stability. Extending this analysis to antimicrobial peptides (AMP), we show that the motifs identified by COLOR have a $>$50\% higher chance of converting a non-AMP sequence to an AMP when used for mutation, compared to those identified by an attention-based XAI method. This demonstrates COLOR's ability to elucidate the role of individual monomers and identify critical motifs in protein sequences. Additionally, COLOR's monomeric contribution score-driven sequence optimization offers a promising alternative to deep generative models, which typically demand large training data. This capability provides the foundation for monomeric contribution score-driven sequence optimization to accelerate the design of de novo proteins that have great potential to be used for nanomedicine, catalysis, and sustainable protein-based biomaterials. 



\section{Data availability}
The authors did not generate any new data for the current study. 

\section{Code availability}
The codes and files necessary files will be shared after the peer-review process. 

\begin{acknowledgement}
The authors acknowledge funding from the National Science Foundation’s MRSEC program (DMR-2308691) at the Materials Research Center of Northwestern University. AP also acknowledges Payal Mohapatra from the Department of Electrical and Computer Engineering at Northwestern University for her valuable input regarding the preparation of this manuscript. 

\end{acknowledgement}

\section{Authors Contribution}
A.P., S.K., and W.C. conceived the idea. A.P. performed all implementations. A.P., S.K., and W.C. contributed to the manuscript writing.

\section{Declaration of Interests}
The authors declare no competing interests.
\bibliography{achemso-demo}
\newpage


\section*{Supplementary material}

# Supplementary information for COLOR: A compositional linear operation-based representation of protein sequences for identification of monomer contributions to properties

Akash Pandey,[1] Wei Chen,[1] and Sinan Keten[*,1,2]

[1]Department of Mechanical Engineering, Northwestern University, Evanston, Illinois 60208, United States

[2]Department of Civil and Environmental Engineering, Northwestern University, Evanston, Illinois 60208, United States

*E-mail: s-keten@northwestern.edu

## 1   Related interpretable models for protein sequences

With advancements in AI, models such as CNN, LSTM, and Transformers have been extensively used to predict properties based on the primary sequence. However, these models lack interpretability due to the layers of added non-linearity. But recently there have been some developments in the field of Explainable AI (XAI) methods to improve the interpretability of DL models. Based on the techniques discussed in the comprehensive review of the XAI

method for biological application by Karim et al.[1*], we are going to use three methods as the baseline due to their applicability to the sequence-based models. These methods are Grad-CAM,[2,3] Attention Tracing,[4-6] and Grad-SAM.[7] The grad-SAM method has not been used for proteins but is a simple extension of the Attention Tracing method; hence we have included it as our baseline. The description of the baseline methods is as follows:

**Grad-CAM**:

The Gradient-weighting Class Activation Mapping (Grad-CAM)[8] was initially proposed for producing a visual explanation for the decision made by a CNN-based model for classification tasks. To understand the formulation of Grad-CAM let us consider $z_{ji}$ (where $i$=1,...L, and $j$=1,...d) to be the latent space representation of the primary sequence with L and d being the sequence length and latent vector size for every position in the sequence respectively. Also, let $y^p$ be the output predicted by the DL model. Then the importance of each position in the primary sequence can be obtained using Supplementary Equation.1.

$$\overline{\phi_i} = \left| \frac{1}{d} \sum_j \frac{\partial y^p}{\partial z_{ji}} \right|$$

(Supplementary Equation.1)

**Attention Tracing**:

Self-attention mechanism in Transformers[9] has been used to study the importance of different positions in the field of Natural Language Processing (NLP),[9] images,[11] and protein.[4] Self-attention captures the impact of one time point (or pixel in the case of images) on another in the sequential data. Transformer consists of several layers and in each layer, self-attention is calculated. Let us indicate the layer number using $n$ and the self-attention matrix of $n^{th}$ layer as $\alpha^{(n)}$. $\alpha^{(n)}$ is calculated as per Supplementary Equation.2a using the Query (Q) and Key(K) matrix obtained from the primary sequence as indicated in Vaswani et al.[9] with $\alpha_{i \to j}^{(n)}$ indicating the attention position $i$ places on $j$ in $n^{th}$ layer of the transformer. Using $\alpha^{(n)}$, the importance $\overline{\phi_i^{(n)}}$ (where i= 1,...L) can be obtained for every position in the sequence

---

*Reference in the Supplementary information (SI) are separate from those in the Main text and are provided at the end of SI.

in every layer of the Transformer with Supplementary Equation.2b.

$$\alpha^{(n)} = \text{softmax}\left(\frac{Q^{(n)}K^{(n)^T}}{d_{K^{(n)}}^{0.5}}\right), \text{ where } \alpha^{(n)} \in R^{L*L} \qquad \text{(Supplementary Equation.2a)}$$

$$\overline{\phi_i^{(n)}} = \sum_{j=1}^{L} \alpha_{i \to j}^{(n)} \phi_j^{(n+1)} \qquad \text{(Supplementary Equation.2b)}$$

Using the same Supplementary Equation.2b, the importance can be propagated from the topmost layer of the transformer to the input of the transformer. The importance of the input to the transformer $(\overline{\phi_i^{(1)}})$ indicates the importance of positions in the primary sequence.

**Grad-SAM**:

Gradient Self-Attention Map (Grad-SAM)[7] has been employed in NLP to identify the input elements that explain the model prediction. Grad-SAM extends the attention tracing method by incorporating an additional gradient term into Supplementary Equation.2b, yielding the formulation in Supplementary Equation.3.

$$\overline{\phi_i^{(n)}} = \sum_{j=1}^{L} \alpha_{i \to j}^{(n)} \phi_j^{(n+1)} \left| \frac{\partial y_p}{\partial \alpha_{i \to j}^{(n)}} \right| \qquad \text{(Supplementary Equation.3)}$$

Including the gradient term helps capture not only the absolute value of the self-attention $(\alpha_{i \to j})$ but also the sensitivity of the output $(y_p)$ with respect to these self-attention values.

## 2    Model details

Supplementary Table 1: Details for our model for every dataset used in the study.

| Dataset | $q$ | $m$ | $d$ | Trainable Parameters |
|---|---|---|---|---|
| Toy Dataset 1 | 5 | 4 | 8 | 10,601 |
| Toy Dataset 2 | 9 | [4,8,6,3] | 8 | 48,657 |
| ACP Aliphatic | 20 | 1 | 20 | 7,341 |
| ACP GRAVY | 20 | 1 | 8 | 7,337 |
| ACP Instability | 20 | 3 | 20 | 11,965 |
| Collagen Melting temperature | 21 | [4,8,6,3] | 8 | 52,433 |
| GB1 binding affinity | 20 | 4 | 20 | 37237 |
| Soluprot | 20 | [4,8,16,24] | 20 | 166,922 |
| AMP classification | 4 | 8 | 4 | 18,958 |
| Silk | 20 | [4,8,6,3] | 8 | 51,089 |

Supplementary Table 2: Details of state-of-the-art models used for comparison with our model. For the silk dataset, there is no data available for 1D CNN and LSTM as this dataset was only used for the explainability study.

| | Models | | |
|---|---|---|---|
| Dataset | Transformer | 1D CNN | LSTM |
| Toy Dataset 1 | 159K | 137K | 115K |
| Toy Dataset 2 | 138K | 138K | 115K |
| ACP Aliphatic | 139K | 87K | 139K |
| ACP GRAVY | 139K | 87K | 139K |
| ACP Instability | 158K | 139K | 139K |
| Collagen Melting temperature | 158K | 139K | 139K |
| GB1 binding affinity | 158K | 139K | 139K |
| Soluprot | 158K | 139K | 139K |
| Silk | 158K | - | - |
| Antimicrobial Classification | 159K | 138K | 163K |

## 3    Additional Analysis

We have designed these toy datasets such that the output properties ($y$) for each primary sequence are determined through an analytical formulation, providing an exact ground truth for model evaluation. This approach is particularly useful for conducting monomeric contribution studies, where the primary sequence is masked based on its contribution score, and

the properties are re-evaluated as described in Sec. **Quantifying Explainability** in the main text. In the case of the toy dataset, re-evaluating the properties is straightforward due to the availability of an analytical formulation.

## 3.1 Supplementary Datasets

### 3.1.1 Toy Dataset 1

In this dataset, the primary sequence consists of 5 qualitative (i.e., categorical) variables: A, B, C, D, and E. At each $i^{th}$ position, the $\psi_i$ is assigned to account for 2 neighboring positions in the sequence as:

$$\psi_i = (a_{i-1} + a_i) * a_{i+1} \qquad \text{(Supplementary Equation.4)}$$

, where, $\psi_0$ and $\psi_L$ are equal to 0. The term $a_i$ takes on values 5,2,4,1, or 8 depending on whether A, B, C, D, or E is present at that position. The property $y$ is calculated as

$$y = \sum_{i=1}^{L} \psi_i \qquad \text{(Supplementary Equation.5)}$$

The $L$ of all the primary sequences is 50.

### 3.1.2 Toy Dataset 2

In this dataset, the primary sequence consists of 9 distinct qualitative variables: A, B, C, D, E, F, G, H, and I. At each $i^{th}$ position, the descriptor $a_i$ is assigned, where $a_i$ takes on values 5, 2, 4, 1, 8, 10, 7, 6, or 3 depending on whether A, B, C, D, E, F, G, H, or I is present at that position. The descriptor at each position is further refined to $\psi_i$ to incorporate the influence of neighboring variables in the sequence as:

$$\psi_i = \sum_{j=i-b_i/2}^{i+b_i/2} a_j \qquad \text{(Supplementary Equation.6)}$$

5

, where, $b_i$ takes values 10, 12, 14, 16, 18, 20, 22, 24, or 26, depending on whether A, B, C, D, E, F, G, H, or I is present at the $i^{th}$ position. The property $y$ is calculated as:

$$y = \sum_{i=1}^{L} \psi_i \qquad \text{(Supplementary Equation.7)}$$

The $L$ of all the primary sequences is 50.

### 3.1.3  Spider silk peak force

Silk has superior mechanical properties, making it a good choice for designing biomaterials. Kim et al.[12] collected the peak force data using Molecular Dynamics (MD) simulation for 82 different primary sequences of silk mimicking MaSp1 spidroin of the spider silk.[13] We use this computational data as one of the datasets for our study. For simplicity, we will refer to this dataset as the "Silk dataset".

The data split for the above-mentioned datasets is given in Supplementary Table.3.

Supplementary Table 3: Data split for different datasets used in the current study.

| Dataset | Training | Validation | Test |
|---|---|---|---|
| Toy Dataset 1,2 | 1000 | 100 | 100 |
| ACP Aliphatic, GRAVY, Instability | 850 | 150 | 150 |
| Silk | 50 | 15 | 15 |

## 3.2   Comprehensive Predictive Capability

To study the predictive capability of COLOR, we use $\mathscr{A}$ and $\mathscr{A}_{500}$, as introduced in Sec.**Quantifying Explainability** in the Main text. The $\mathscr{A}$ and $\mathscr{A}_{500}$ values for Toy dataset 1 and 2 are plotted in Supplementary Figure.1 along with all other datasets discussed in the Main text. On the toy datasets, COLOR outperforms the next-best SOTA model by 34% on average. We do not perform this study on the silk dataset due to the insufficiency of training samples to calculate either $\mathscr{A}$ or $\mathscr{A}_{500}$. An important observation from Supplementary Figure.1 is the strong performance of COLOR on Toy Dataset 2. In this dataset, the monomeric properties

$(\psi)$ depend on a larger neighborhood of monomers, as defined by Supplementary Equation.4. Interestingly, despite using small values of $m$ (3,4,6,8) in the model, COLOR accurately predicts the property $y$ as evident by the lower values of $\mathscr{A}$ and $\mathscr{A}_{500}$. This highlights the model's ability to capture properties at the motif level through $\mathbf{Q}$, while the matrix multiplication of $\mathbf{D}$ and $\mathbf{Q}^T$, combined with the non-linearities in the neural network, enables the model to effectively capture global interactions.

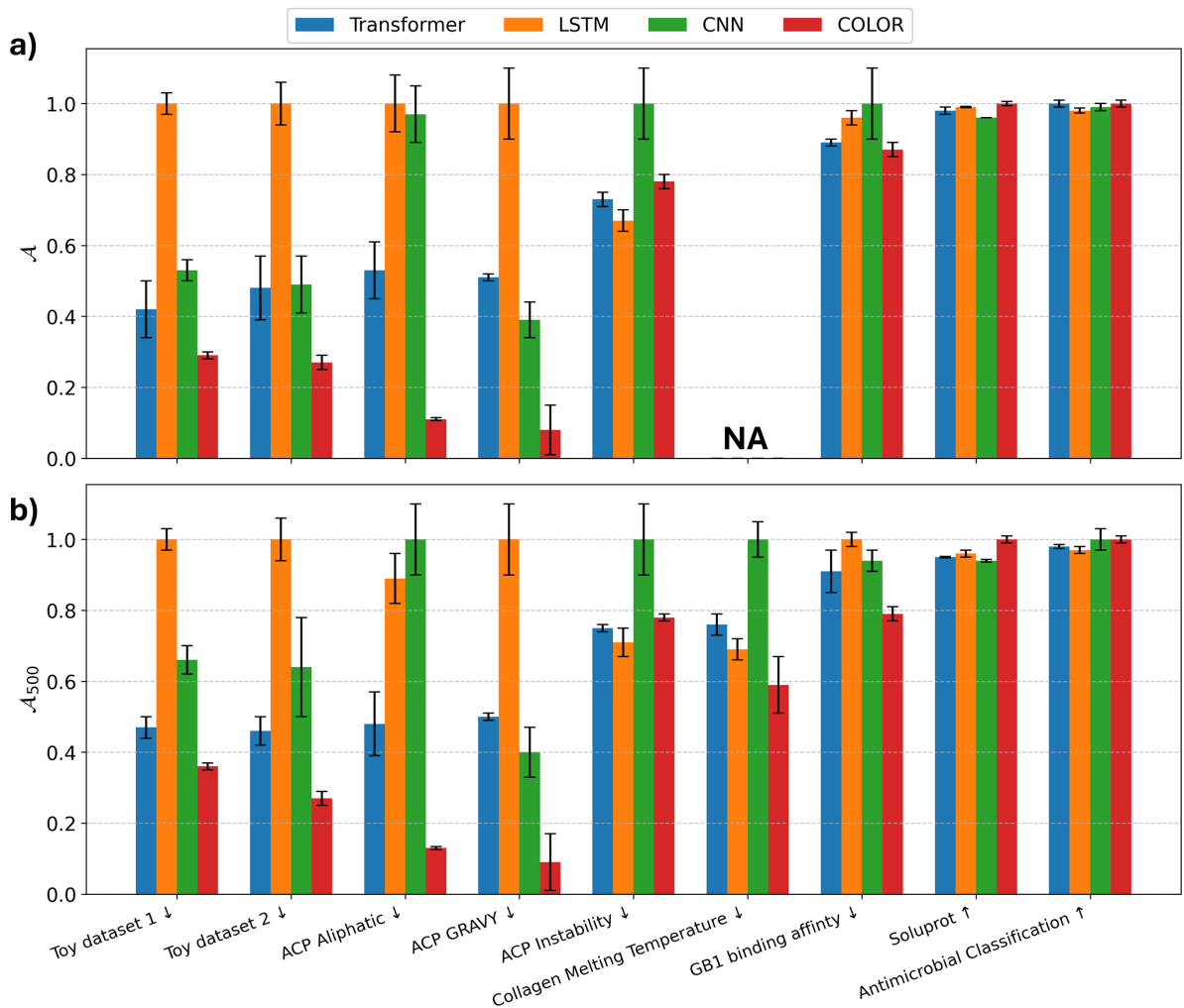

Supplementary Figure. 1: Comparison of the predictive capability of different supervised models. Figure a) shows the comparison of $\mathscr{A}$ and b) illustrates the comparison of $\mathscr{A}_{500}$ obtained for different datasets. The arrows ↑ and ↓ in front of dataset names indicate whether higher or lower values are better, respectively. The results in the table are *normalized* using the highest values of the corresponding dataset.

## 3.3 Comprehensive Explainability

We study the explainability of the COLOR method in estimating the monomeric contribution scores for two toy datasets and the silk dataset. Even though we do not perform the predictive study on the Silk dataset, we include it in the explainability study. Using the data split provided in Supplementary Table.3 for the Silk dataset, the mean $R^2$ value of 0.91 is achieved across all models, indicating strong predictive performance and making the dataset suitable for the explainability study. We use the metric $\mathscr{I}$ discussed in Sec.**Monomeric Contribution Calculation Method** in the Main text to quantify the explainability. The values of $\mathscr{I}$ are shown in the Supplementary Figure.2 along with all the datasets discussed in the Main text. The COLOR method outperforms the next-best method by 37%, 5%, and 14% on Toy Dataset 1, Toy Dataset 2, and the Silk dataset, respectively.

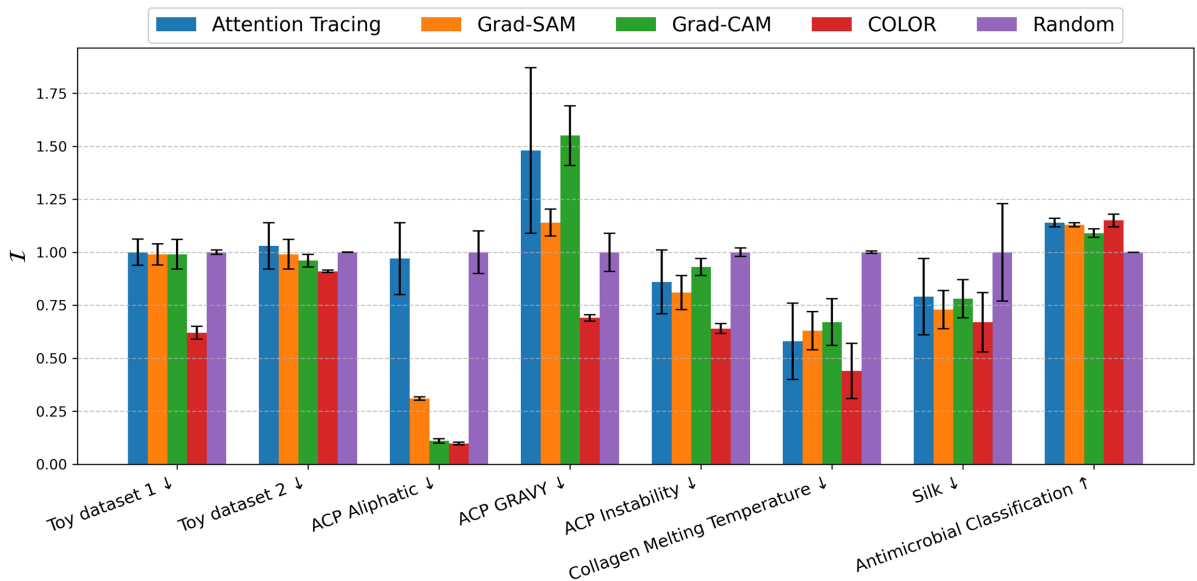

Supplementary Figure. 2: Comparison of explainability offered by different XAI models. The results are normalized using the results from the *Random method* of the corresponding dataset. The arrows ↑ and ↓ in front of dataset names indicate whether higher or lower values are better, respectively.

# 4    Motif Identification in a Toy Dataset

We show the capability of COLOR to effectively identify important motifs using Toy Dataset 1 as the critical motif is correctly known in this case. Supplementary Figure.3 showcases the capability of Grad-SAM and COLOR methods in accurately pinpointing the critical motifs within the sequences. The Grad-SAM method is chosen for the comparison as it ranks the second-best interpretable model for toy dataset 1 as shown in Supplementary Figure.2. Based on the results in Supplementary Figure.3, COLOR successfully identifies the most important motif in 8 out of 10 instances, compared to 6 out of 10 for the Grad-SAM method.

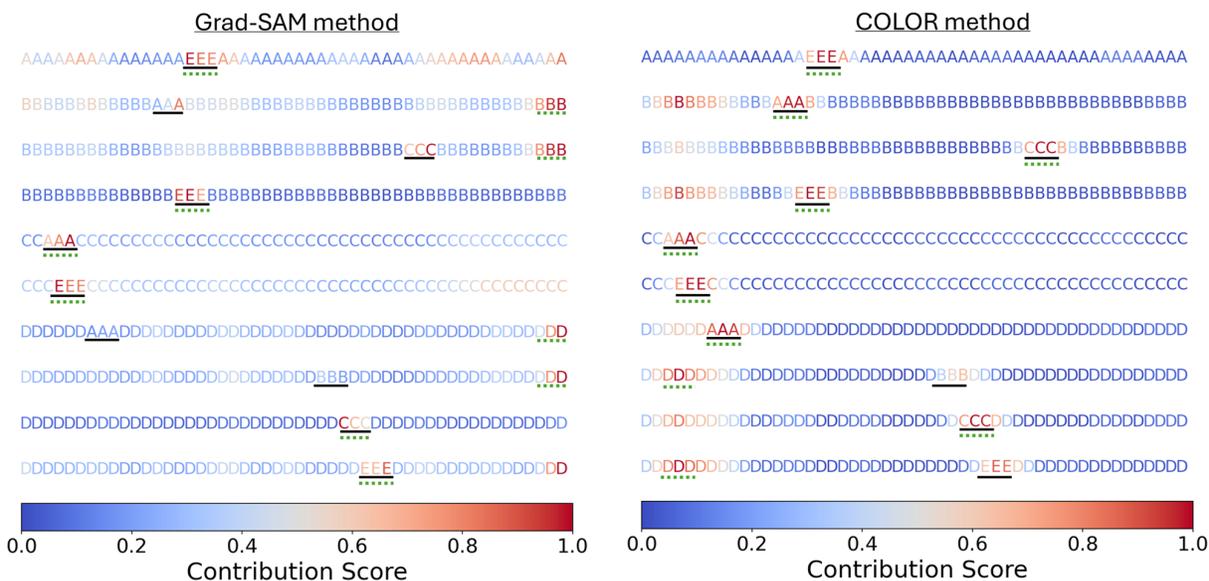

Supplementary Figure. 3: Motif identification in sequences from Toy dataset 1. The figure highlights the most contributing motif in each sequence with a black underline. Every monomer in the sequence is color-coded based on its contribution score, as determined using the Grad-SAM and COLOR methods. Based on the contribution scores, the key motif identified by Grad-SAM and COLOR methods is shown using green underline. With a good interpretable model, the black solid and green dashed underlines are expected to overlap more frequently.

# 5 Strategies for Enhancing Model Performance

In Sec. **Complete Architecture**, we discuss the overall architecture of our model. Based on the architecture, we outline several strategies to enhance model performance across various applications:

- Adjust the motif size $m$ to identify the optimal choice for different datasets, as they may contain motifs of varying sizes.

- Tune the latent space dimension $d$ to balance representation capacity and model complexity.

- Increase the number of COLOR units, each with unique motif sizes $m$ to capture diverse motif structures.

- Apply layer normalization after each COLOR unit for enhanced optimization in certain applications.

- Introduce layer normalization following the 3D representation $\mathbf{R}$ for further optimization benefits.

- Normalize the motif composition matrix $\mathbf{D}$ by the sequence length to ensure the model processes relative compositions rather than absolute values. This approach is particularly advantageous for properties like the Aliphatic Index, which depends on the relative proportions of amino acids A, V, I, and L.

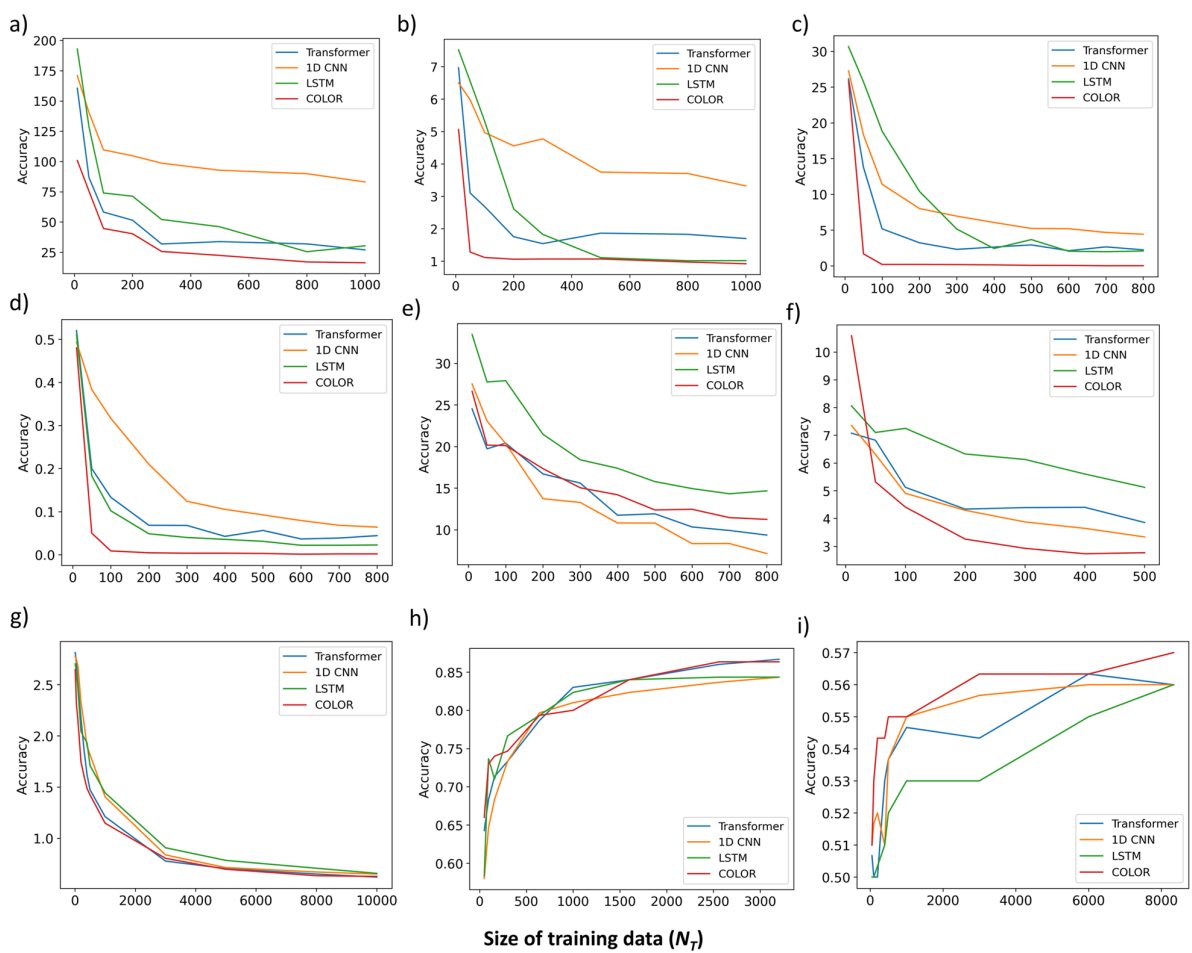

Supplementary Figure. 4: Curves showing the predictive capability of different models. MAE (or accuracy) versus training data size ($N_T$) curves obtained from different deep-learning models are plotted for a) Toy dataset 1, b) Toy dataset 2, c) ACP Aliphatic index dataset, d) ACP GRAVY index dataset, e) ACP instability dataset, f) Collagen dataset, g) Silk dataset, h) Antimicrobial classification dataset. and i) Soluprot dataset. The curves shown here are the mean of runs using three different seeds.

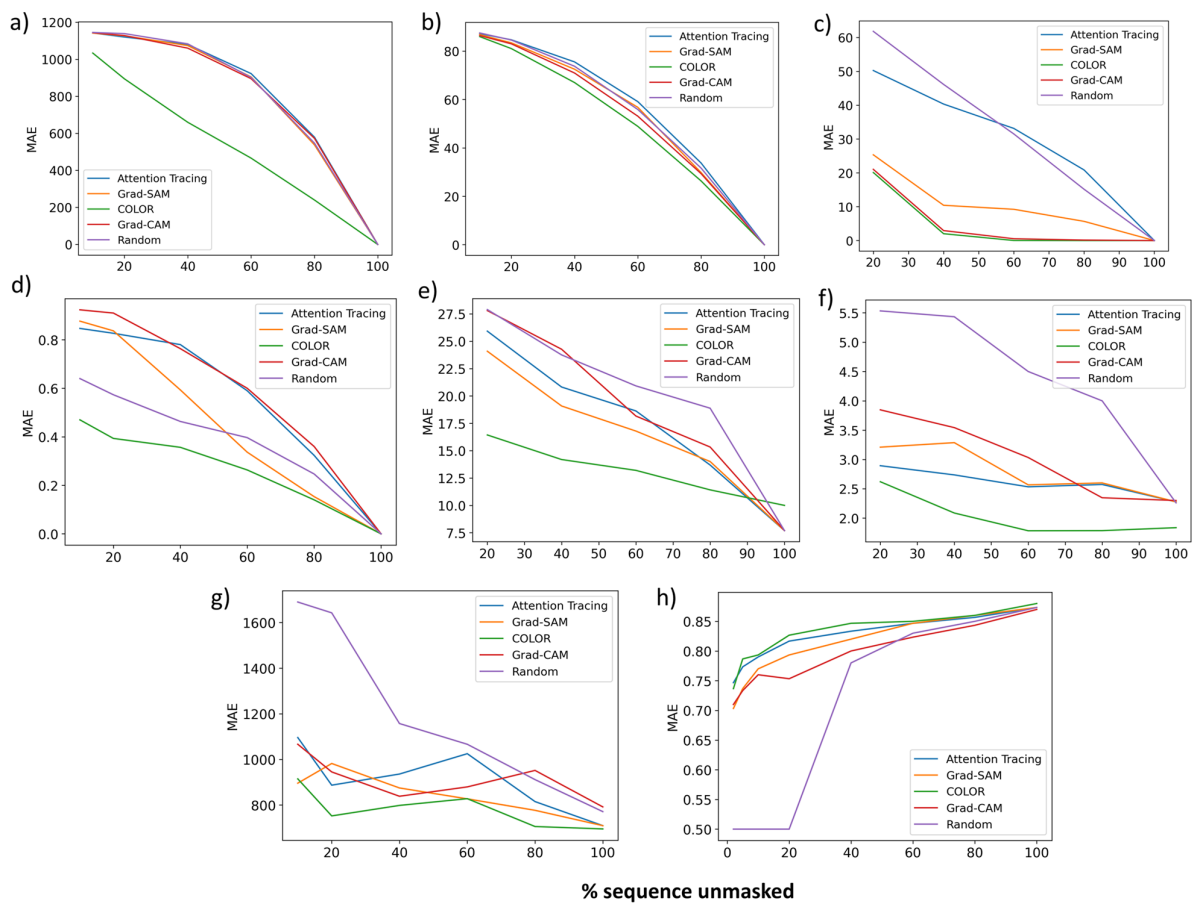

Supplementary Figure. 5: Curves showing the explainability of different models. MAE (or accuracy) versus % of sequence unmasked curve obtained from different XAI models are plotted for a) Toy dataset 1, b) Toy dataset 2, c) ACP Aliphatic index dataset, d) ACP GRAVY index dataset, e) ACP instability dataset, f) Collagen dataset, g) Silk dataset, and h) Antimicrobial classification dataset. The curves shown here are the mean of runs using three different seeds.

# Supplementary References


\end{document}